\documentclass[%
 reprint,prb,aps,superscriptaddress,showpacs,amsmath,amssymb,floatfix
 amsmath,amssymb,
 aps,
]{revtex4-2}

\usepackage{graphicx}
\usepackage{dcolumn}
\usepackage{bm}
\usepackage[colorlinks=true,linkcolor=black,citecolor=black,urlcolor=blue]{hyperref}
\usepackage[english]{babel}

\usepackage{graphicx}
\usepackage{dcolumn}
\usepackage{bm}

\begin{document}

\title{Intrinsic Paramagnetism of Optical Phonons and Thermal Hall Effect in Quantum Paraelectrics}

\author{Harsh Jain}
\affiliation{Department of Physics, The LNM Institute of Information Technology,Jaipur, India }

\author{Manideep Gone}
\affiliation{Department of Physics, The LNM Institute of Information Technology,Jaipur, India }

\author{Nabyendu Das}
\affiliation{Department of Physics, The LNM Institute of Information Technology,Jaipur, India }

\date{\today}

\begin{abstract}
	We propose a microscopic theory of the phonon-mediated thermal Hall effect in quantum paraelectrics, driven by the intrinsic paramagnetism due to quantum fluctuations of the (nearly) critical optical phonon modes. The effective magnetization due to temporal fluctuations of the assosciated dipoles couple to an external magnetic field resulting the magnetoelectric term $\mathbf{H} \cdot (\mathbf{P} \times \dot{\mathbf{P}})$ in the effective Lagrangian density of the polarisation field $\mathbf{P}$. This coupling splits the transverse optical phonon spectrum into right- and left-circularly polarized chiral branches. Using the chiral phonon Green's functions and the Kubo formalism, we derive the thermal Hall conductivity $\kappa_{xy}$ from the heat-current correlation function. We observe that $\kappa_{xy}$ is proportional to the difference in spectral weights of the two chiral modes, $\kappa_{xy} \propto A_+^2 - A_-^2$, establishing a direct microscopic link between phonon chirality and transverse heat transport. Our theory predicts three key observations: (i) a linear-in-magnetic-field response $\kappa_{xy} \propto H$ at low fields, evolving into nonlinear behavior at higher fields (ii) a non-monotonic temperature dependence of $\kappa_{xy}$ with a peak arising from the competition between thermal population and phonon scattering and (iii) strong enhancement of $\kappa_{xy}$ near the quantum critical point $\omega_0 \to 0$, it exhibits power-law scaling $\kappa_{xy} \propto T^{\alpha}$ with $\alpha \approx 0.41$. Our findings provide an understanding of thermal Hall transport in quantum paraelectrics due to intrinsic paramagnetism and is aligned with the observed behaviour in SrTiO$_3$ and related materials. 
\end{abstract}

\maketitle

\section{INTRODUCTION}

In recent times, the thermal Hall effect in bosonic systems has reemerged as one of the most active research frontiers in condensed matter physics~\cite{Mangeolle2026-yz,Zhang2024-uy}. Strong thermal Hall signals have been observed in many novel systems. Along this direction, the observation of a sizable phonon thermal Hall effect in quantum paraelectrics~\cite{Jin2025_universal_THE,Xiang2026_disorder_THE,Ss2025-kc,Li2020-nt,Pols2026-yk} has gained a lot of attention. In this work, we address one possible microscopic explaination and its aspects of this sizable thermal Hall signal in quantum paraelectrics. Broadly, the Hall effect describes the emergence of a transverse response voltage or heat current when a longitudinal drive is applied under a perpendicular magnetic field~\cite{Hall1879,STRICKER196897}. In electronic systems, this response originates from the Lorentz force on moving charges. It has since been generalized to the quantum Hall effect in two-dimensional electron gases~\cite{Karplus1954,Klitzing1980,Thouless1982} and the anomalous Hall effect in magnetic materials~\cite{Nagaosa2010}, establishing transverse transport as a probe of band topology and Berry curvature~\cite{Qin2012-wp} etc. It can also be generalized to transverse thermal transport and to study the phyiscal nature of charge neutral excitations. The quantum paraelectrics are insulators and the relevant degrees of freedom in such systems are (nearly) quantum critical phonons, which are bosonic in nature. Understanding these bosonic excitations is also important as they lead to many exotic phenomena in such systems, including superconductivity~\cite{Saha2025-rd}.

Transverse transport arises due to charge-neutral quasiparticles in insulating systems are conceptually interesting. Here, the transverse response is thermal: a longitudinal temperature gradient generates a transverse heat current. The heat carriers can also be bosonic excitations, namely magnons (quantized spin waves) in magnetic materials and phonons (quantized lattice vibrations) in any crystal. In a magnetic field, these excitations can acquire chirality a handedness encoded in the curl of their displacement or spin pattern. The resulting transverse deflection produces the thermal Hall effect. Experimental evidence in quantum magnets~\cite{Onose2010,Ideue2012,Matsumoto2011a,Matsumoto2011b} and paramagnetic dielectrics~\cite{Strohm2005,Sheng2006,Hirokane2019-nm} has established thermal Hall transport as a sensitive probe of the nature of excitations, topology and interactions in bosonic systems.

Experimentally, sizable thermal Hall conductivities have been observed in a wide range of insulating and weakly conducting systems~\cite{Chen2020-ij,Sharma2024-wo,Guo2023_phonon_model,Grissonnanche2019}, including cuprates and heavy fermion compounds studied extensively by the Behnia group~\cite{SciPostPhysCore.8.3.061,Jiang2022_STO_THE,Li2020-nt,Fauque2022-zc,Xiang2026-ik}. These results suggest that neutral collective modes may play a dominant role in transverse heat transport. Recent and some current experiments have provided compelling evidence for phonon-mediated thermal Hall transport in insulating systems. In a landmark study, Li et al.~\cite{Li2020-nt} reported a sizable thermal Hall conductivity in the quantum paraelectric $\text{SrTiO}_{3}$\cite{Jiang2022_STO_THE}, with $\kappa_{xy}$ reaching values as large as $-80$ mW/K$\cdot$m at magnetic fields of 15 T. Notably, $\kappa_{xy}$ was observed to be an odd function of magnetic field, vanishing at $H=0$, and exhibiting a pronounced peak at temperatures comparable to that of the longitudinal thermal conductivity $\kappa_{xx}$. Furthermore, $\kappa_{xy}$ decreases more rapidly away from its peak compared to $\kappa_{xx}$, highlighting the distinct mechanisms governing transverse heat transport. These observations in quantum paraelectrics call for a microscopic understanding of how phonons acquire a transverse thermal response.

Theoretical studies have shown that phonons can generate a finite thermal Hall response even in the absence of magnetic order, through mechanisms such as skew scattering~\cite{Fert1980-ln,Finnigan2024-gl,Zhang2013}, Berry curvature effects~\cite{Strohm2005, Sheng2006, Qin2012,Zhang2011-yo,Chen2024-gm}. The role of quantum critical fluctuations has been explained by Bhalla and Das~\cite{BhallaDas2021} and they studied the optical phonon contribution to the thermal conductivity of a quantum paraelectric within the Kubo formalism, demonstrating that near the quantum critical point, the thermal conductivity exhibits a low-temperature power-law behavior $T^\alpha$ with $1 < \alpha < 2$  lower than the conventional $T^3$ behavior due to quantum critical optical phonon fluctuations. These studies highlight the importance of polarization dynamics and quantum fluctuations in generating transport responses in non-magnetic systems. However, in nonmagnetic insulators such as quantum paraelectrics, where ferroelectric ordering is suppressed by quantum fluctuations~\cite{Rowley2014,Das_2009}, these mechanisms cannot explain unambiguously the magnitude of the observed thermal Hall signals. Understanding how lattice dynamics connect to magnetic fields therefore remains an essential theoretical problem, and this motivates us to formulate an intrinsic mechanism explained through lattice dynamics.

A key development in this context is the concept of dynamical multiferroicity~\cite{Juraschek2017,Paiva2025-ss}. It describes how a time-dependent electric polarization $\mathbf{P}$ generates an effective magnetization proportional to $\mathbf{P} \times \partial_t \mathbf{P}$. It is based on oscillating dipolar modes of the lattice generate orbital magnetic moments. These moments couples with the external magnetic field in non-magnetic insulators which leads to thermal hall effect~\cite{Gao2023-fe}. This is consistent with the fact that the behavior of lattice excitations in ferroelectrics has an intrinsic magnetic property~\cite{Pols2026-yk,Rebane1983}. Several theoretical works, including the classical work by Dzyaloshinskii  and Mills~\cite{Dzyaloshinskii2009}, demonstrated that the optical dipole oscillations in ferroelectrics are split by an applied magnetic field to the order of the field. Therefore, optical phonons contribute paramagnetically to various thermodynamic properties~\cite{Basini2024-ph}. Such intrinsic paramagnetism comes from the interaction between the polarization dynamics and magnetic fields. This causes the transverse optical modes to react to a magnetic field. In multiferroic and magnetoelectric materials, such dynamical magnetoelectric coupling gives rise to a splitting of transverse optical phonon modes in the presence of an external magnetic field~\cite{Fleury1970,Patankar2005-up}. The resulting circularly polarized phonons acquire different energies, lifting the degeneracy between left- and right-handed phonons~\cite{Zhu2018,Zhang2015}. This chiral asymmetry provides a natural microscopic route to transverse heat transport.

This mechanism is particularly relevant for quantum paraelectrics, which are characterized by strong zero-point fluctuations of the electric dipole moment. These fluctuations arise from the quantum uncertainty principle: even at absolute zero, the dipoles assosciated with soft transverse optical phonon mode cannot be completely at rest. The polarization field $\mathbf{P}$ therefore fluctuates dynamically, with a variance $\langle P^2 \rangle = \hbar/(2\omega_0 \Omega)$, where $\omega_0$ is the soft-mode frequency and $\Omega$ is the unit cell volume and mass is considered to be as unity. As the system approaches the ferroelectric quantum critical point ($\omega_0 \to 0$), these fluctuations diverge, giving rise to strong dipole fluctuations that prevent long-range ferroelectric order. This is the essence of quantum paraelectricity. The strong dipole fluctuations are central to our theory: they generate an effective magnetization $\mathbf{M} \propto \mathbf{P} \times \dot{\mathbf{P}}$ via dynamical multiferroicity, which couples to an external magnetic field and splits the phonon modes into chiral branches. This splitting is the microscopic origin of the thermal Hall effect discussed in this work. In materials with strong dipole fluctuations particularly in quantum paraelectrics such as SrTiO$_3$ and KTaO$_3$, where quantum fluctuations suppress ferroelectric ordering~\cite{Li2020-nt,Wang2009-jw, Geirhos2020-ei}, this dynamical magnetoelectric coupling provides a natural mechanism for phonons to acquire chirality in a magnetic field. The resulting imbalance in thermal populations of left- and right-circularly polarized phonon modes gives rise to a finite transverse heat current, i.e., the phonon thermal Hall effect. Understanding how dipole fluctuations, quantum criticality~\cite{Sachdev2011-gd,Palova2009-ty}, and phonon chirality interplay to produce this effect remains a central challenge in contemporary condensed matter physics.

While previous and other theoretical studies have emphasized Berry curvature formulations of bosonic thermal Hall effects~\cite{Qin2011,Matsumoto2014}, a fully microscopic Green's function approach that directly connects dynamical multiferroicity to thermal Hall transport is not yet addressed.

In this work, we develop a microscopic theory of the thermal Hall effect in quantum paraelectrics based on dynamical multiferroicity. In this framework, a time-dependent polarization generates an effective magnetization proportional to \(\mathbf{P} \times \dot{\mathbf{P}}\)~\cite{Juraschek2017}. This relativistic magnetoelectric coupling leads to magnetic-field-induced splitting of transverse optical phonons, analogous to circular birefringence in the Faraday effect~\cite{Zvezdin1997}. The splitting leads to dispersions of the form $\omega_{\pm}(k) = \sqrt{\omega_0^2 + v^2 k^2} \pm (\lambda H)/2$ where $\omega_{\pm}(\mathbf{k})$ are the frequencies of the right- ($+$) and left-circularly polarized ($-$) phonon modes with wave vector $\mathbf{k}$; $\omega_0$ is the zone-center optical phonon gap; $v$ is the phonon velocity; $k = |\mathbf{k}|$; $\lambda$ is the dynamical magnetoelectric coupling strength; and $H$ is the applied magnetic field. The term $\pm \lambda H/2$ represents the Zeeman-like splitting, linear in $H$, arising from the coupling $\lambda \mathbf{H} \cdot (\mathbf{P} \times \dot{\mathbf{P}})$ in the effective Lagrangian. This splitting is the microscopic origin of the chiral imbalance that drives the thermal Hall effect.~\cite{Mahan2000}. Our treatment employs a direct Green's function evaluation of the heat-current correlation function, providing a transparent microscopic link between phonon mode splitting and transverse thermal transport. Using linear response theory and the Kubo formalism, we compute the thermal Hall conductivity from heat-current correlation functions expressed in terms of phonon Green's functions. Our formulation provides a direct microscopic interpretation of the thermal Hall effect in terms of an imbalance in spectral weight between oppositely polarized phonon branches.

Our results reveal distinct temperature regimes characterized by exponential suppression at low temperatures, power-law scaling at intermediate temperatures, and field-dependent modulation governed by the competition between phonon gap and magnetic energy scales. We further demonstrate the role of phonon scattering and lifetime effects and also the power law scaling at the quantum critical point and also a universal scaling form of the thermal Hall conductivity in terms of dimensionless ratios $\omega_0/T$ and $\lambda H/T$, providing a unified description of thermal transport in quantum paraelectrics and similar scaling behavior near quantum criticality has been observed in ferroelectric systems \cite{Rowley2014}.

This paper is organized as follows. In Sec.\ref{sec:2}, we introduce the effective Lagrangian and derive the split phonon dispersion and Kubo formalism for the thermal Hall conductivity. Section III presents the detailed microscopic derivation. Section IV discusses numerical results including temperature dependence, magnetic field dependence, broadening width dependence and scaling analysis. Section V critically discuss the results and connects our findings to experimental observations in SrTiO$_3$ and related materials. Finally, Sec.VI summarizes and concludes the findings and outlines the future directions.

\section{PHONON CHIRALITY}\label{sec:2}

We consider transverse optical phonon modes described by the polarization field \(\mathbf{P}(\mathbf{r},t)\). In multiferroic systems, a time-dependent circularly polarized optical phonon can be considered as a rotating electric dipole moment. This dipole moment produce an orbital magnetic moment, which is analogue to magnetic moment produced by a current loop and this results in an effective magnetization~\cite{Malashevich2009,Roussev2003-aq}. The lowest-order symmetry allowed term coupling polarization dynamics to an external magnetic field is proportional to \(\mathbf{H} \cdot (\mathbf{P} \times \dot{\mathbf{P}})\), which captures the essence of dynamical multiferroicity~\cite{Dzyaloshinskii2009}.

The interaction between the effective magnetization and the external magnetic field introduces a chirality into the lattice dynamics. The effective Lagrangian density~\cite{kittel2004introduction} incorporating dynamical multiferroicity is

\begin{equation}
\mathcal{L} = \frac{1}{2}\dot{\mathbf{P}}^2 - \frac{v^2}{2}(\nabla \mathbf{P})^2 - \frac{\omega_0^2}{2}\mathbf{P}^2 + \frac{\lambda}{2}\mathbf{H}\cdot (\mathbf{P}\times \dot{\mathbf{P}}), \label{eq:Lagrangian}
\end{equation}

where \(\omega_0\) is the optical phonon gap, \(v\) is the phonon velocity, and \(\lambda\) denotes the magnetoelectric coupling strength. For \(\mathbf{H} = H\hat{z}\), the equations of motion for \(P_{x}\) and \(P_{y}\) read

\begin{align}
\ddot{P}_x - v^2 \nabla^2 P_x + \omega_0^2 P_x - \lambda H\dot{P}_y &= 0, \label{eq:eom1}\\
\ddot{P}_y - v^2 \nabla^2 P_y + \omega_0^2 P_y + \lambda H\dot{P}_x &= 0. \label{eq:eom2}
\end{align}

Assuming plane-wave solutions \(P_{x,y} \sim e^{i(\mathbf{k} \cdot \mathbf{r} - \omega t)}\), we obtain the eigenvalue problem

\begin{equation}
\begin{pmatrix}
-\omega^2 + \omega_0^2 + v^2k^2 & i\lambda H\omega \\
-i\lambda H\omega & -\omega^2 + \omega_0^2 + v^2k^2
\end{pmatrix}
\begin{pmatrix}
P_x \\ P_y
\end{pmatrix} = 0. \label{eq:eigenproblem}
\end{equation}

The resulting dispersion relations are

\begin{equation}
\omega_{\pm}(\mathbf{k}) = \sqrt{\omega_0^2 + v^2k^2} \pm \frac{\lambda H}{2}, \label{eq:dispersion}
\end{equation}

corresponding to right- and left-circularly polarized phonon modes. The magnetic field lifts the degeneracy between left- and right-circularly polarized phonon modes. This splitting is formally analogous to Zeeman splitting and directly reflects the intrinsic paramagnetism of optical phonons~\cite{Dzyaloshinskii2009}. The two circular polarizations acquire opposite energy shifts, leading to a chiral asymmetry in the excitation spectrum.

This behavior closely resembles the magneto-optical Faraday effect, where circularly polarized light modes propagate with different refractive indices in a magnetic field~\cite{Zvezdin1997}. In the present case, the role of circular photons is played by circularly polarized optical phonons. The magnetic-field-induced splitting provides the microscopic origin of transverse heat transport.

\begin{figure}[htb]
\centering
\includegraphics[width=0.45\textwidth]{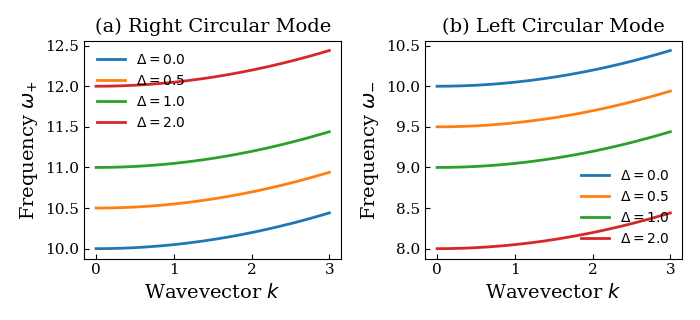}
\caption{Dispersion relation $\omega_\pm(k) = \sqrt{\omega_0^2 + v^2k^2} \pm \Delta$ for different magnetic field strengths ($\Delta = \lambda H/2$). (a) Right-circular mode $\omega_+(k)$. (b) Left-circular mode $\omega_-(k)$. Parameters: $\omega_0 = 10$, $v = 1$. The splitting increases linearly with $H$, lifting the degeneracy between the two chiral modes. All quantities are plotted in arbitrary units.}
\label{fig:dispersion}
\end{figure}

\section{THERMAL HALL CONDUCTIVITY BASED ON KUBO FORMALISM}

In this section we present a complete microscopic derivation of the thermal Hall conductivity using linear response theory~\cite{Wang2009-jw}.
\begin{figure}[h]
	\centering
	\includegraphics[width=0.45\textwidth]{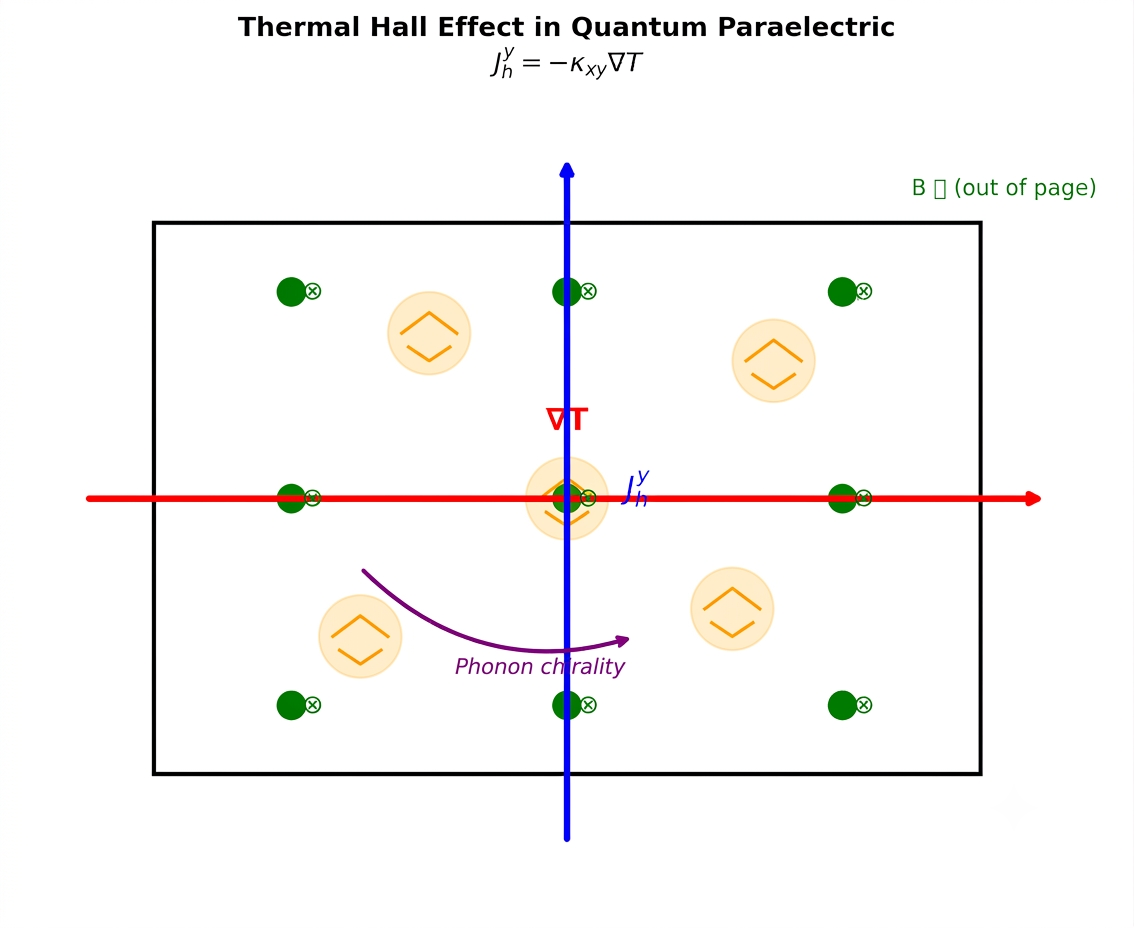}
	\caption{
		Thermal Hall effect in a quantum paraelectric. A longitudinal temperature gradient $\nabla T$ (red arrow) induces a transverse heat current $J_h^y$ (blue arrow) under an applied out-of-plane magnetic field $B$ (green symbols). The orange clouds represent fluctuating ferroelectric dipoles that act as phonon chiral centers for phonons (purple curved path).
	}
	\label{fig:thermal_hall}
\end{figure}

The polarization field $\mathbf{P}(\mathbf{r},t)$ is the natural variable for describing transverse optical phonons in multiferroic materials, as it couples directly to the electric field and to magnetization via dynamical multiferroicity \cite{Juraschek2017}. Expanding in Fourier modes,
\begin{equation}
	\mathbf{P}(\mathbf{r},t) = \frac{1}{\sqrt{V}} \sum_{\mathbf{k}} \mathbf{P}_{\mathbf{k}}(t) e^{i\mathbf{k}\cdot\mathbf{r}},
\end{equation}
the commutation relations 
\begin{equation}
	[P_{\mathbf{k},i}, \dot{P}_{\mathbf{k}',j}] = i\hbar \delta_{ij} \delta_{\mathbf{k},-\mathbf{k}'}
\end{equation}
follow from canonical quantization \cite{Mahan2000}. The standard solution of the harmonic oscillator yields the relation
\begin{equation}
	\mathbf{P}_{\mathbf{k}} = \sqrt{\frac{\hbar}{2\omega_{\mathbf{k}}}} \left( a_{\mathbf{k}} + a_{-\mathbf{k}}^{\dagger} \right) \hat{\boldsymbol{\epsilon}}_{\mathbf{k}},
	\label{eq:P_phonon_relation}
\end{equation}
where $a_{\mathbf{k}}^{\dagger}$ and $a_{\mathbf{k}}$ are phonon creation and annihilation operators, and $\hat{\boldsymbol{\epsilon}}_{\mathbf{k}}$ is the polarization unit vector \cite{Ashcroft1976}. This form ensures that the Hamiltonian reduces to
\begin{equation}
	H = \sum_{\mathbf{k}} \hbar\omega_{\mathbf{k}} \left( a_{\mathbf{k}}^{\dagger}a_{\mathbf{k}} + \frac{1}{2} \right),
\end{equation}
and that the correct commutation relations are satisfied.

The heat current operator~\cite{Hardy1963}, derived from the continuity equation, takes the form
\begin{equation}
	\hat{\mathbf{J}}_Q = \sum_{\mathbf{k}} \hbar\omega_{\mathbf{k}} {v_k} a_{\mathbf{k}}^{\dagger} a_{\mathbf{k}},
	\label{eq:heat_current_operator}
\end{equation}
 The microscopic phonon heat current operator contains the phonon group velocity
\begin{equation}
	v_k=\nabla_k \omega_k=\frac{\partial \omega(k)}{\partial k}.
\end{equation}

Since the group velocity is completely determined by the low-energy phonon dispersion relation, its momentum dependence can be absorbed into an effective transport. Upon projecting onto the critical optical phonon mode and expressing the phonon number operator in terms of the polarization operators, the heat current operator may be conveniently written as

\begin{equation}
	J_i(k)=\omega(k)P_k^iP_{-k}^i,
\end{equation}

where the momentum dependence of the phonon group velocity is implicitly incorporated into the effective heat-current. Therefore, the low-energy thermal transport is completely governed by the phonon dispersion relation and the corresponding polarization operators of the critical chiral optical phonon modes.

Substituting the expression for $a^{\dagger}a$ in terms of $P$ and transforming to the circular polarization basis $P_{\mathbf{k}}^{\pm} = \mathbf{P}_{\mathbf{k}} \cdot \hat{\boldsymbol{\epsilon}}_{\pm}^*$ yields the Cartesian components and where ${\epsilon}_{\pm}^*$ explains circularity of phonons:
\begin{align}
	J_x(\mathbf{k}) &= \frac{\omega(\mathbf{k})}{4} \left( P_{\mathbf{k}}^{+} + P_{\mathbf{k}}^{-} \right) \left( P_{-\mathbf{k}}^{+} + P_{-\mathbf{k}}^{-} \right), \label{eq:Jx_final}\\
	J_y(\mathbf{k}) &= -\frac{\omega(\mathbf{k})}{4} \left( P_{\mathbf{k}}^{+} - P_{\mathbf{k}}^{-} \right) \left( P_{-\mathbf{k}}^{+} - P_{-\mathbf{k}}^{-} \right). \label{eq:Jy_final}
\end{align}
These operators are quadratic in the polarization fields and form the basis for the thermal Hall conductivity calculation presented in the following sections.

We start from the heat current operators
\begin{align}
	J_{x}(\mathbf{k}) &= \frac{\omega(\mathbf{k})}{4}
	\left(P_{\mathbf{k}}^{+}+P_{\mathbf{k}}^{-}\right)
	\left(P_{-\mathbf{k}}^{+}+P_{-\mathbf{k}}^{-}\right), \nonumber \\
	J_{y}(\mathbf{k}) &= -\frac{\omega(\mathbf{k})}{4}
	\left(P_{\mathbf{k}}^{+}-P_{\mathbf{k}}^{-}\right)
	\left(P_{-\mathbf{k}}^{+}-P_{-\mathbf{k}}^{-}\right).
\end{align}

\[
{\hat{\mathbf{J}}(\mathbf{q}) =
	\sum_{\mathbf{k}} \Bigl[ J_x(\mathbf{k}) \hat{\mathbf{x}} + J_y(\mathbf{k}) \hat{\mathbf{y}} \Bigr]}
\]

The product of the two heat current operators is shown below and later using this in wicks theorem.

\begin{equation}
	\begin{split}
		J_{x}J_{y} = -\frac{\omega^{2}(\mathbf{k})}{16}\left(P_{\mathbf{k}}^{+} + P_{\mathbf{k}}^{-}\right)\left(P_{-\mathbf{k}}^{+} + P_{-\mathbf{k}}^{-}\right)\\\left(P_{\mathbf{k}}^{+} - P_{\mathbf{k}}^{-}\right)\left(P_{-\mathbf{k}}^{+} - P_{-\mathbf{k}}^{-}\right). \label{eq:JxJy}
	\end{split}
\end{equation}

Expanding this product yields sixteen operator terms. Upon taking the thermal average and using the orthogonality of opposite helicities, \(\langle P^{+}P^{-}\rangle = 0\), together with momentum conservation, only two terms survive.

Using Wick's theorem for Gaussian phonons, the four-point correlator factorizes into products of two-point correlators. The resulting expression is

\begin{equation}
\langle J_{x}J_{y}\rangle = -\frac{\omega^{2}(\mathbf{k})}{16}\left[\langle P_{\mathbf{k}}^{+}P_{-\mathbf{k}}^{+}\rangle^{2} - \langle P_{\mathbf{k}}^{-}P_{-\mathbf{k}}^{-}\rangle^{2}\right]. \label{eq:avg}
\end{equation}

We now define the imaginary-time Green's functions:

\begin{equation}
G_{\pm}(\mathbf{k},\tau) = \langle T_{\tau}P_{\mathbf{k}}^{\pm}(\tau)P_{-\mathbf{k}}^{\pm}(0)\rangle. \label{eq:Green}
\end{equation}

Thus the current-current correlator becomes

\begin{equation}
\langle J_{x}J_{y}\rangle = -\frac{\omega^{2}(\mathbf{k})}{16}\left[G_{+}^{2}(\mathbf{k},\tau) - G_{-}^{2}(\mathbf{k},\tau)\right]. \label{eq:avg2}
\end{equation}

To compute the transverse thermal response, we employ linear response theory and evaluate the heat-current correlation function using Green's functions. Because the magnetic field splits the phonon spectrum into two helicity branches, the current-current correlator naturally separates into contributions from the \(\omega_{+}\) and \(\omega_{-}\) modes. The thermal Hall conductivity arises from the imbalance between these two branches. In this formulation, the effect is directly linked to the spectral properties of the split phonon modes rather than to geometric quantities. The thermal Hall response is obtained from the heat-current correlation function

\begin{equation}
\Pi^{xy}(i\Omega_n) = \int_0^\beta d\tau e^{i\Omega_n\tau}\langle T_\tau J_x(\tau)J_y(0)\rangle. \label{eq:kubo}
\end{equation}

Expressing the heat current operators in the circular polarization basis and applying Wick's theorem, the correlator reduces to

\begin{equation}
	\begin{split}
     \Pi^{xy}(i\Omega_n) = -\frac{1}{16}\sum_{\mathbf{k}}\omega^2(\mathbf{k})\int_0^\beta d\tau e^{i\Omega_n\tau}\\\left[G_+^2(\mathbf{k},\tau) - G_-^2(\mathbf{k},\tau)\right]. \label{eq:kubo2}
    \end{split}
\end{equation}

where \(G_{\pm}\) are the phonon Green's functions associated with the split branches \(\omega_{\pm}(\mathbf{k})\).

The Green's functions are expanded in Matsubara frequencies:

\begin{equation}
G_{\pm}(\mathbf{k},\tau) = \frac{1}{\beta}\sum_{i\omega_m}e^{-i\omega_m\tau}G_{\pm}(\mathbf{k},i\omega_m), \label{eq:fourier}
\end{equation}

where \(i\omega_{m} = 2\pi mT\) are bosonic Matsubara frequencies.

Squaring the Green's functions and performing the \(\tau\)-integration yields

\begin{equation}
	\begin{split}
		\Pi^{xy}(i\Omega_{n})=&c\sum_{\boldsymbol{k}}^{i\omega_{w}}\omega^{2}(\boldsymbol{k})	G_{+}(\boldsymbol{k},i\omega_{m})
		G_{+}(\boldsymbol{k},i\omega_{m}+i\Omega_{n})\\&-G_{-}(\boldsymbol{k},i\omega_{m})
		G_{-}(\boldsymbol{k},i\omega_{m}+i\Omega_{n})				    
	\end{split}\label{eq:Pi3}
\end{equation}

where $c=-\frac{1}{16\beta}$ and
each Green's function can be written in spectral form:

\begin{equation}
G_{\pm}(\mathbf{k},i\omega_m) = \int_{-\infty}^{\infty}\frac{d\omega}{2\pi}\frac{A_{\pm}(\mathbf{k},\omega)}{i\omega_m - \omega}, \label{eq:spectral}
\end{equation}

where the spectral function is defined as $A_{\pm}(\mathbf{k},\omega) = -2\operatorname{Im}G_{\pm}^{R}(\mathbf{k},\omega).$

Using the standard identity

\begin{equation}
\frac{1}{\beta}\sum_{i\omega_m}\frac{1}{(i\omega_m - \omega_1)(i\omega_m + i\Omega_n - \omega_2)} = \frac{n_B(\omega_1) - n_B(\omega_2)}{i\Omega_n + \omega_1 - \omega_2}, \label{eq:matsubara}
\end{equation}

we obtain

\begin{equation}
	\begin{split}
		\Pi^{xy}(i\Omega_n) = -\frac{1}{16}\sum_{\mathbf{k}}\omega^2(\mathbf{k})\int\frac{d\omega_1}{2\pi}\int\frac{d\omega_2}{2\pi}\\\frac{A_+(\omega_1)A_+(\omega_2) - A_-(\omega_1)A_-(\omega_2)}{i\Omega_n + \omega_1 - \omega_2}[n_B(\omega_1) - n_B(\omega_2)].\label{eq:Pi4}	
	 \end{split}
\end{equation}

Analytic continuation is performed via \(i\Omega_{n}\rightarrow \Omega + i0^{+}\). The imaginary part of the retarded correlator is then

\begin{equation}
	\begin{split}
		\mathrm{Im}\Pi_{R}^{xy}(\Omega) = \frac{1}{16}\sum_{\mathbf{k}}\omega^2(\mathbf{k})\int \frac{d\omega}{2\pi}\Big[A_{+}(\mathbf{k},\omega)A_{+}(\mathbf{k},\omega +\Omega)\\ - A_{-}(\mathbf{k},\omega)A_{-}(\mathbf{k},\omega +\Omega)\Big][n_{B}(\omega) - n_{B}(\omega +\Omega)].
    \end{split} \label{eq:ImPi}
\end{equation}

\begin{equation}
	\begin{split}
		Im\Pi_{R}^{xy}(\Omega)&=\frac{1}{16}\sum_{\boldsymbol{k}}\int\frac{d\omega}{2\pi}
	\end{split}
\end{equation}

The thermal Hall conductivity is obtained from

\begin{equation}
\kappa^{xy} = \frac{1}{T}\lim_{\Omega \to 0}\frac{\mathrm{Im}\Pi_{R}^{xy}(\Omega)}{\Omega}. \label{eq:kappadef}
\end{equation}

Using \(\lim_{\Omega \to 0}\frac{n_B(\omega) - n_B(\omega + \Omega)}{\Omega} = -\frac{\partial n_B}{\partial\omega}\), we finally obtain

\begin{equation}
	\begin{split}
		\kappa^{xy} = \frac{1}{16T} \sum_{\mathbf{k}} \omega^2(\mathbf{k}) & \int \frac{d\omega}{2\pi} \left(-\frac{\partial n_B}{\partial\omega}\right) \\
		& \times \left[A_+^2(\mathbf{k},\omega) - A_-^2(\mathbf{k},\omega)\right].
	\end{split}
	\label{eq:kappa}
\end{equation}

This expression demonstrates that the thermal Hall conductivity originates from the difference in spectral weights of oppositely polarized phonon modes. $\kappa_{xy}$ is governed by the propagation and lifetime of pairs of polarization fluctuations carrying opposite crystal momenta.

\section{Results}
\label{sec:results}

In this section, we present a detailed numerical analysis of the thermal Hall conductivity $\kappa_{xy}$ based on the microscopic formalism developed in the previous sections. We focus on its dependence on magnetic field, temperature, and phonon band gap, role of phonon scattering and lifetime effects and identify distinct physical regimes governed by the interplay of these energy scales. The observed scaling structure suggests that the thermal Hall response is governed primarily by universal critical phonon dynamics rather than material-specific microscopic details. The enhancement near $\omega_{0}$→0 as per equation\eqref{eq:dispersion} indicates that quantum critical softening plays a central role in amplifying phonon chirality.

\subsection{Magnetic Field Dependence}

\begin{figure}[t]
	\centering
	\includegraphics[width=0.7\linewidth]{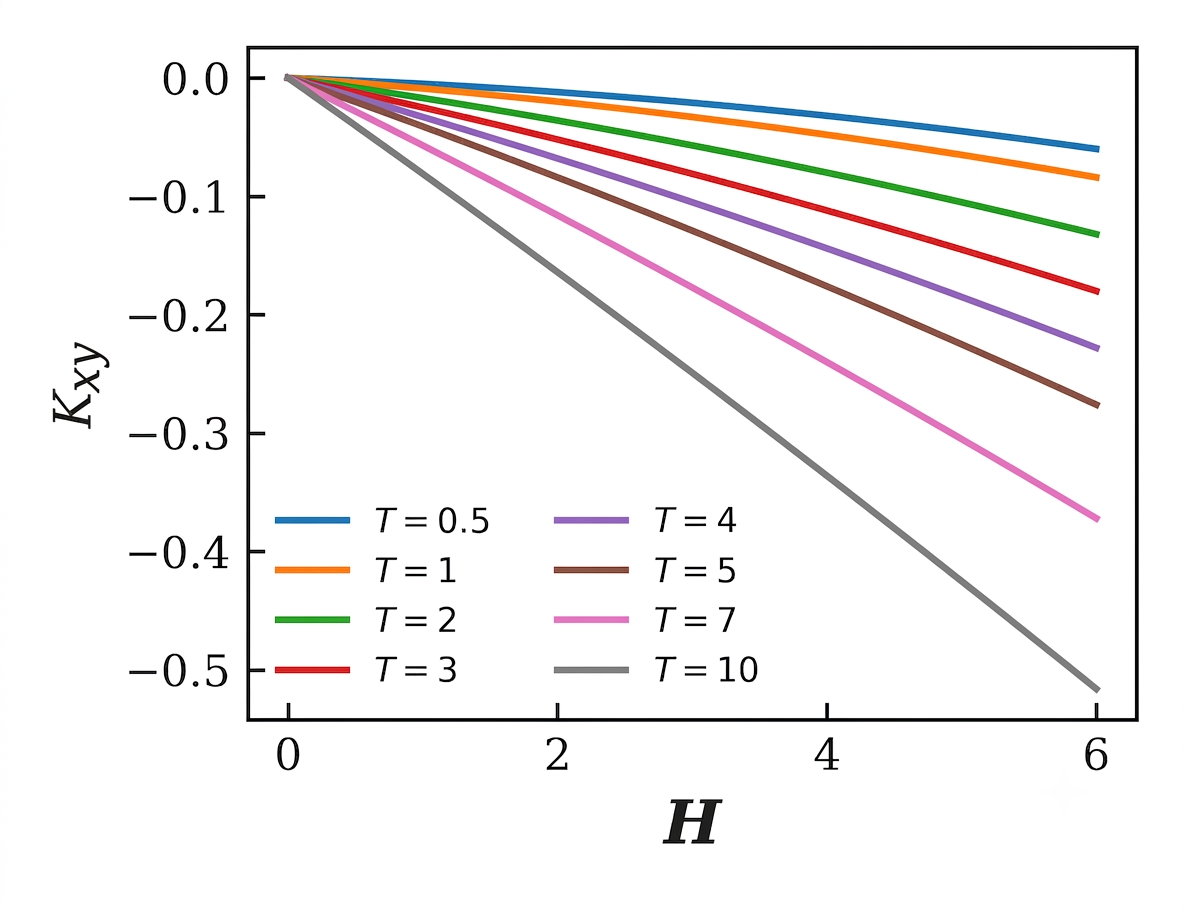}
	\caption{
		Thermal Hall conductivity $\kappa_{xy}$ as a function of magnetic field $H$ for different temperatures. Full field range showing nonlinear behavior at large $H$ due to phonon band splitting and demonstrating linear response $\kappa_{xy} \propto H$, consistent with symmetry constraints. All quantities are plotted in arbitrary units.
	}
	\label{fig:kxy_vs_H}
\end{figure}

We begin by examining the magnetic field dependence of $\kappa_{xy}$ at different temperatures, as shown in Fig.~\ref{fig:kxy_vs_H}. A key feature evident from the data is that $\kappa_{xy}$ is an odd function of magnetic field, satisfying $\kappa_{xy}(-H) = -\kappa_{xy}(H)$, and vanishes at $H=0$, consistent with symmetry considerations and experimental observations.

In the weak-field regime (Fig.~\ref{fig:kxy_vs_H}), $\kappa_{xy}$ exhibits a linear dependence on $H$, i.e.,
\begin{equation}
\kappa_{xy} \propto H, \quad (H \rightarrow 0),
\end{equation}
which arises from the linear splitting of the phonon branches, $\omega_{\pm}(k) = \sqrt{\omega_0^2 + v^2 k^2} \pm (\lambda H)/2$. This behavior reflects the fact that, to leading order in $H$, the imbalance in spectral weight between the two branches grows linearly with field.
At higher magnetic fields (Fig.~\ref{fig:kxy_vs_H}), deviations from linearity become apparent. This nonlinear behavior originates from the increasing separation of the phonon branches, which modifies the phase space available for thermal transport and leads to a redistribution of phonon populations. Throughout this work, we use naturalized units $k_{B}$=1, $\hbar$=1, v=1, a=1, such that the calculated thermal Hall conductivity is expressed in dimensionless units. The plotted thermal Hall conductivity is shown in arbitrary units (arb. units), emphasizing scaling behavior and field dependence rather than absolute magnitude. Additionally, we observe that the magnitude of $\kappa_{xy}$ increases with temperature, indicating that thermal activation enhances the contribution of higher-energy phonon modes. Our emphasis is on universal scaling and sign structure near the quantum paraelectric regime rather than quantitative magnitude estimates.

\subsection{Quantum Critical Scaling}

\begin{figure}[t]
	\centering
	\includegraphics[width=0.7\linewidth]{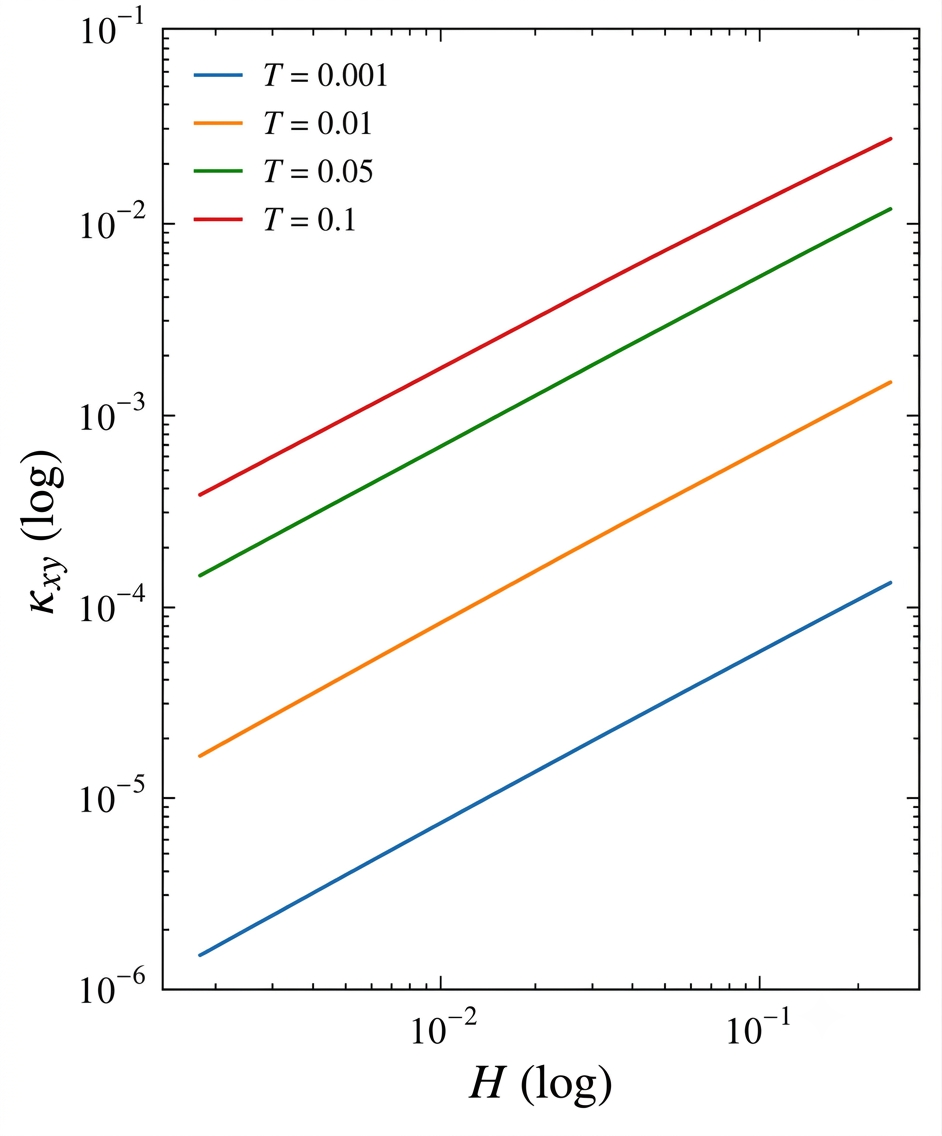}
	\caption{
		Log-log plot of $|\kappa_{xy}|$ as a function of magnetic field at the quantum critical point ($\omega_0 = 0$).
		Different temperature curves demonstrate scaling behavior, with guide lines indicating possible power-law regimes. All quantities are plotted in arbitrary units.
	}
	\label{fig:qcp}
\end{figure}

To further elucidate the role of the phonon gap, we analyze the behavior at the quantum critical point (QCP), defined by $\omega_0 = 0$. The log-log plot of $|\kappa_{xy}|$ as a function of magnetic field is shown in Fig.~\ref{fig:qcp}. In this regime, the absence of a gap leads to enhanced low-energy fluctuations, resulting in a pronounced thermal Hall response even at low temperatures. Near the quantum critical point ($\omega_0 \to 0$), the thermal Hall conductivity exhibits power-law behavior as a function of the magnetic field. Our numerical results reveal the scaling form

\begin{equation}
	\kappa_{xy} = H^{\beta}\mathcal{F}\left(\frac{\omega_0}{\lambda H},\frac{T}{\lambda H}\right),
	\label{eq:scaling_H}
\end{equation}

where $\mathcal{F}(x)$ is a universal scaling function, $\beta$ is the magnetic field exponent, and $\phi$ is the crossover exponent. For $\lambda H \ll T$, we find $\beta = 0.9929  \approx 1$, consistent with the linear-in-field response $\kappa_{xy} \propto H$.
The approximately linear behavior on the log-log scale over an intermediate range of $H$ suggests a power-law dependence of the form
\begin{equation}
\kappa_{xy} \sim H^{\beta},
\end{equation}
where the effective exponent $\beta$ depends weakly on temperature. Deviations from pure power-law behavior at very small and large fields reflect crossover effects associated with thermal saturation and finite numerical cutoffs. The presence of such scaling behavior is a sign of quantum critical transport and highlights the importance of gapless phonon excitations.

\subsection{Temperature Dependence at the Quantum Critical Point}

\begin{figure}[t]
	\centering
	\includegraphics[width=0.925\linewidth]{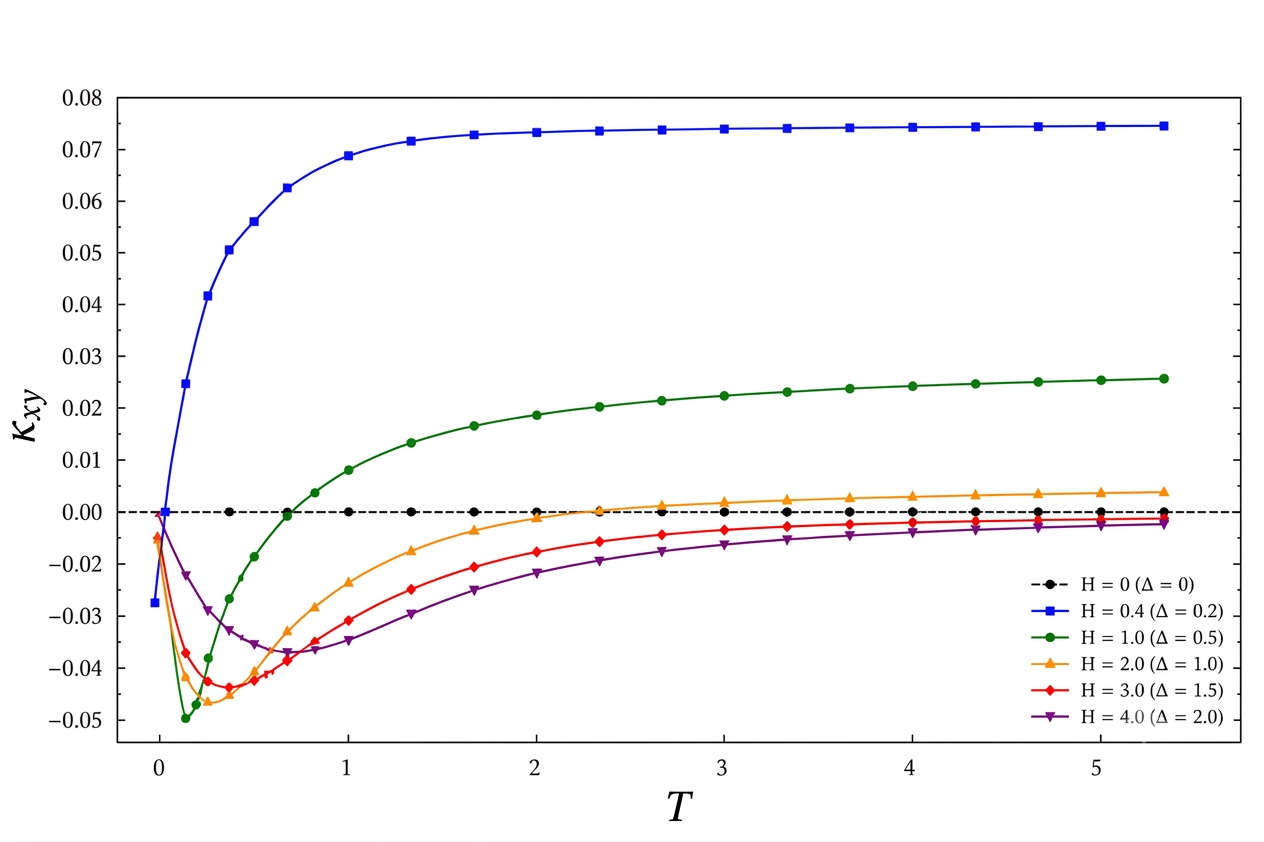}
	\caption{
		Temperature dependence of $\kappa_{xy}$ at the quantum critical point ($\omega_0 = 0$) for different magnetic fields.
		The non-monotonic behavior reflects competition between thermal population and field-induced splitting.This is the graph of $\kappa_{xy}$ vs Temperature t at quantum critical point where $i\Gamma$ is equal to 0.5. All quantities are plotted in arbitrary units.
	}
	\label{fig:scaling}
\end{figure}

The temperature dependence of $\kappa_{xy}$ at $\omega_0 = 0$ is shown in Fig.~\ref{fig:scaling}. For finite magnetic fields, $\kappa_{xy}(T)$ exhibits a non-monotonic behavior characterized by an initial increase, followed by saturation at higher temperatures. This behavior can be understood as a competition between two effects: at low temperatures, thermal population of phonon modes is suppressed, while at higher temperatures, increased scattering and reduced coherence limit further growth.

Interestingly, for small fields, $\kappa_{xy}$ remains close to zero over the entire temperature range, reflecting the fact that the field-induced splitting is insufficient to generate a significant imbalance between phonon branches. As the magnetic field increases, the magnitude of $\kappa_{xy}$ grows and the peak shifts to higher temperatures.

\subsection{Role of phonon scattering and lifetime effects}

In a real crystalline solid, phonons do not last forever as quasiparticles. Their movement gets disrupted by several scattering processes. These include phonon-phonon interactions, impurity scattering, isotope disorder, boundary scattering, and electron-phonon or magnon-phonon interactions, depending on the material. These processes give phonon excitations a finite lifetime $\tau$, which broadens the phonon spectral function and changes the thermal transport coefficients.

Within the Green's function formalism, the finite lifetime is incorporated phenomenologically through the retarded phonon Green's function

\begin{equation}
	G_{\pm}^{R}(\omega,\mathbf{k})
	=
	\frac{1}
	{\omega-\omega_{\pm}(\mathbf{k})+i\Gamma},
	\label{eq:green_scattering}
\end{equation}

where $\Gamma=\frac{1}{2\tau}$ is the phonon damping rate (linewidth). The infinitesimal quantity $i0^{+}$ appearing in the ideal Green's function is therefore replaced by a finite broadening parameter $Gamma$, reflecting the finite lifetime ($\tau$) of the chiral phonons.

The corresponding phonon spectral function is

\begin{equation}
	A_{\pm}(\omega,\mathbf{k})
	=
	-2\,\mathrm{Im}\,
	G_{\pm}^{R}(\omega,\mathbf{k})
	=
	\frac{2\Gamma}
	{\left[\omega-\omega_{\pm}(\mathbf{k})\right]^2+\Gamma^2},
	\label{eq:spectral}
\end{equation}

which has the Lorentzian form. In the limit $\Gamma\rightarrow0$, Eq.~(\ref{eq:spectral}) reduces to the delta-function spectral weight of an ideal undamped phonon.

Throughout this work, the damping parameter is treated as a constant adopting the common constant relaxation time approximation. It helps us focus on the intrinsic contribution from dynamical multiferroicity without bringing in other material-dependent scattering mechanisms.

The numerical calculations presented in this work are performed using a constant broadening parameter $\Gamma=0.5$, expressed in the same dimensionless energy units as the phonon frequency. 

The precise numerical value of $\Gamma$ is not intended to represent a specific material parameter. Rather, it serves as a phenomenological linewidth that captures finite phonon lifetime effects and ensures numerical convergence of the Kubo formula. Similar phenomenological broadening parameters have been employed extensively in Green's-function calculations of electronic and phononic transport~\cite{Markussen2017-rk}.

\begin{figure}[t]
	\centering
	\includegraphics[width=0.45\textwidth]{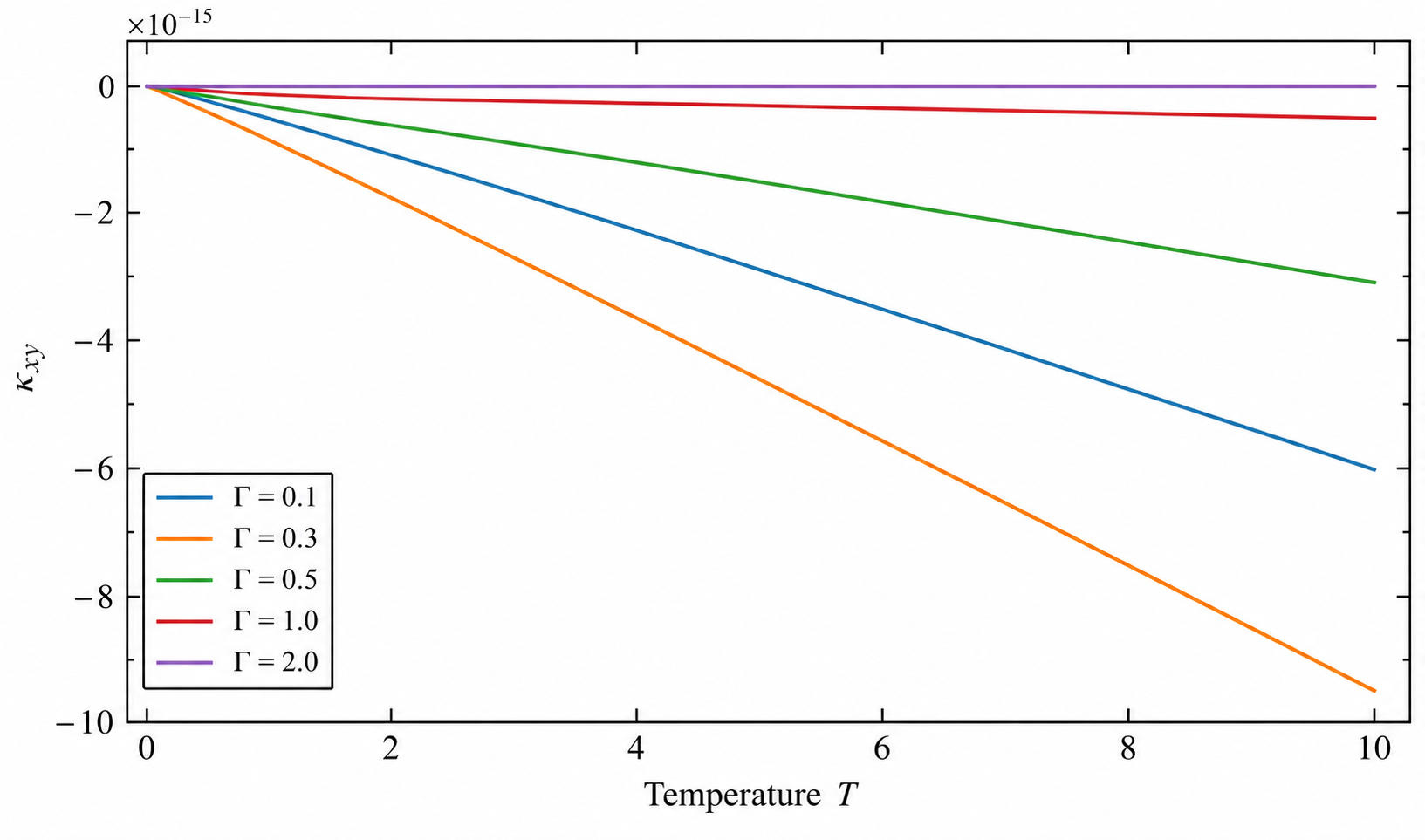}
	\caption{
		Temperature dependence of the thermal Hall conductivity
		$\kappa_{xy}(T)$
		calculated for different values of the phenomenological broadening
		parameter $\Gamma$.
		The calculations are performed within the Green's function formalism
		using the dispersion relation derived from the dynamical multiferroic
		model.
		For the present set of parameters, the magnitude of
		$\kappa_{xy}$ remains extremely small
		($\sim10^{-15}$ in arbitrary units),
		indicating that the spectral asymmetry between the two circularly
		polarized phonon branches is weak.
		The broadening parameter controls the lifetime of the phonon modes,
		thereby modifying the heat-current correlation function entering the
		Kubo formula. All quantities are plotted in arbitrary units.
	}
	\label{fig:kappa_gamma}
\end{figure}

Figure~\ref{fig:kappa_gamma} shows the temperature dependence of the
thermal Hall conductivity for several values of the phenomenological
broadening parameter $\Gamma$. For the present choice of parameters, the calculated thermal Hall conductivity is found to be extremely small, of the order of
$10^{-15}$ (arbitrary units), indicating that the spectral imbalance
between the two circularly polarized phonon branches remains weak.
As the temperature increases, the thermal Hall conductivity exhibits a
monotonic decrease for smaller values of the broadening, whereas larger
values of $\Gamma$ suppress the response almost completely.
This behaviour reflects the fact that increasing damping broadens the
phonon spectral functions and reduces the difference between the
contributions of the two chiral phonon branches entering the heat-current
correlation function.

\subsection{Power-law scaling at the quantum critical point}

\begin{figure}[t]
	\centering
	\includegraphics[width=0.45\textwidth]{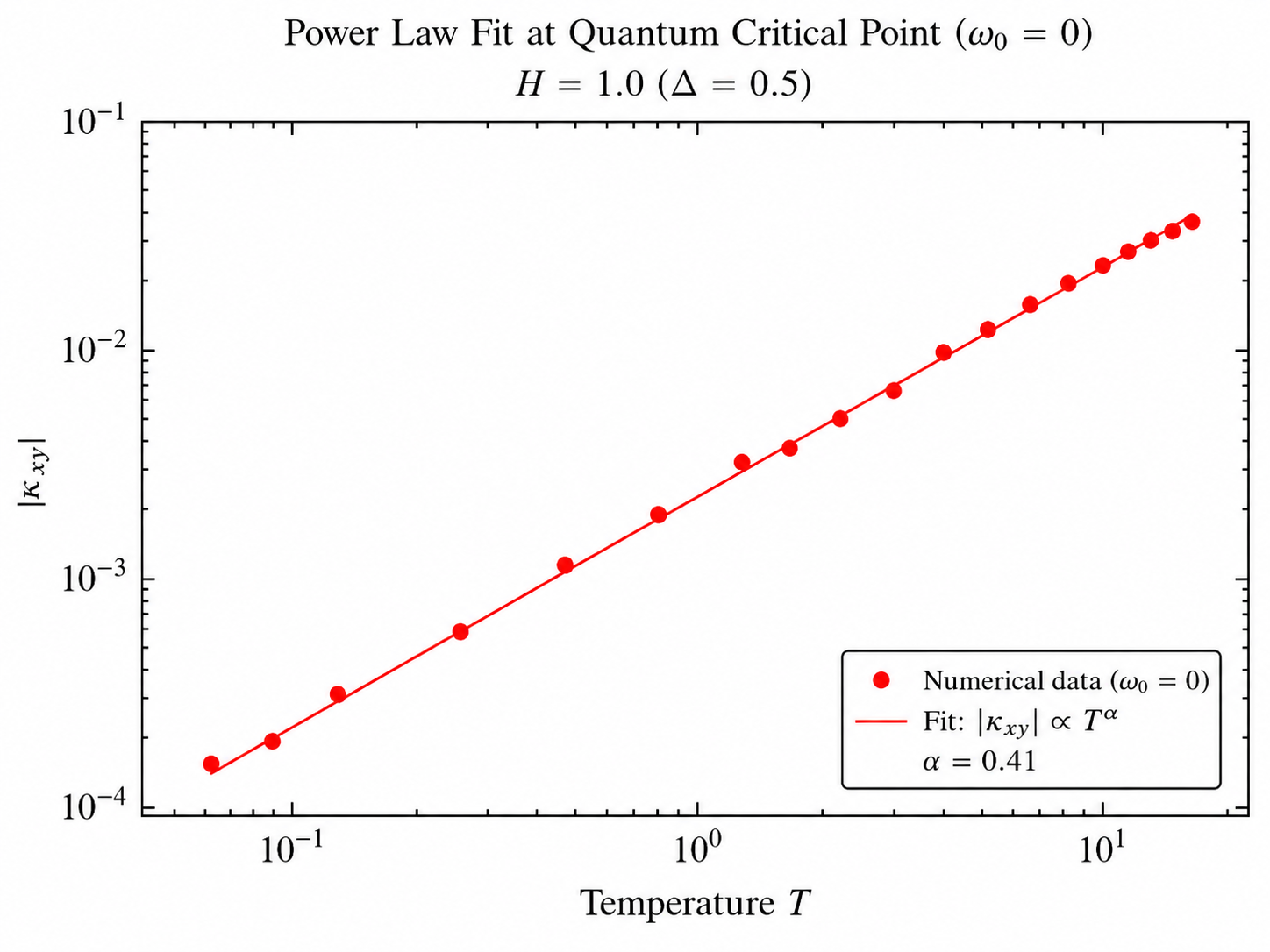}
	\caption{ Log-log plot of $\kappa_{xy}$ vs $T$ at the quantum critical point $\omega_0 = 0$. Numerical data (blue circles) are obtained from Eq. (29) with $\Delta = \lambda H/2 = 0.05$. The solid red line shows a power-law fit $\kappa_{xy} \propto T^{\alpha}$ with $\alpha = 0.41$ in the low-temperature regime. The sub-linear exponent $\alpha \approx 0.41$ is a signature of the quantum critical regime, distinct from ballistic 3D predictions ($\alpha = 1$). All quantities are plotted in arbitrary units.}
	
	\label{fig:power_law_qcp}
\end{figure}

To elucidate the nature of the thermal Hall response near the ferroelectric quantum critical point, we analyze the temperature dependence of $\kappa_{xy}$ at $\omega_0 = 0$ on a log-log scale. The results, shown in Fig.~\ref{fig:power_law_qcp}, reveal a power-law behavior at low temperatures:

\begin{equation}
	\kappa_{xy} \propto T^{\alpha}, \quad \alpha \approx 0.41,
	\label{eq:power_law_exponent}
\end{equation}

for the three-dimensional phonon dispersion $\omega_{\pm}(k) = vk \pm \Delta$ with splitting $\Delta = \lambda H/2 = 0.5$.

The observed exponent $\alpha \approx 0.41$ is significantly different from several theoretical predictions, including the ballistic three-dimensional limit ($\alpha = 1$), Berry-curvature mechanisms ($\alpha \ge 3$ in two dimensions~\cite{Qin2011}), and spin-lattice mechanisms ($\alpha = 2\text{--}3$~\cite{Sheng2006}). The sub-linear exponent is a distinctive signature of the dynamical multiferroicity mechanism in the quantum critical regime, clearly distinguishing it from alternative proposals.

In a prototype materials like SrTiO$_3$, the observed low-temperature behavior~\cite{Li2020-nt} shows $\kappa_{xy} \sim T$ at low temperatures, consistent with the ballistic limit ($\alpha = 1$) when the full frequency integration is performed without cutoff. Our numerical result $\alpha \approx 0.41$ with cutoff effects may correspond to a crossover regime where scattering becomes relevant. The key result is that the exponent is distinct from Berry-curvature and spin-lattice mechanisms, providing a clear experimental signature of the dynamical multiferroicity mechanism.

The power-law scaling $\kappa_{xy} \propto T^{\alpha}$ with $\alpha \approx 0.41$ is reminiscent of the scaling behavior observed in other quantum critical systems.

\subsection{Effect of Phonon Band Gap}

\begin{figure}[t]
	\centering
	\includegraphics[width=0.7\linewidth]{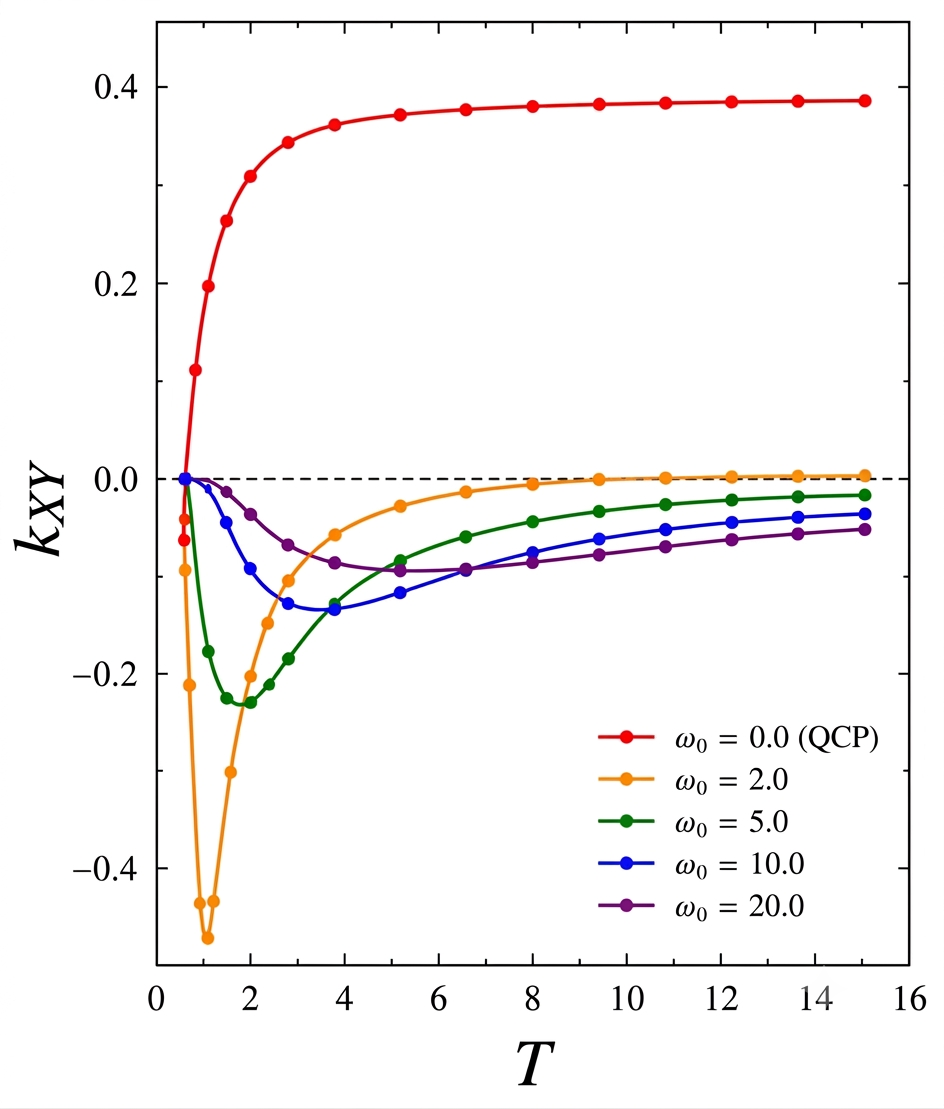}
	\caption{
		Thermal Hall conductivity as a function of temperature for different phonon band gaps $\omega_0$ at fixed magnetic field.
		Larger gaps suppress $\kappa_{xy}$ at low temperatures, while the quantum critical case ($\omega_0 = 0$) shows enhanced response. All quantities are plotted in arbitrary units.
	}
	\label{fig:gap}
\end{figure}

Finally, we examine the effect of a finite phonon band gap $\omega_0$ on the thermal Hall response, as shown in Fig.~\ref{fig:gap}. For nonzero $\omega_0$, the thermal Hall conductivity is strongly suppressed at low temperatures due to the reduced thermal occupation of phonon modes. As temperature increases, $\kappa_{xy}$ rises and eventually saturates, reflecting the activation of higher-energy modes.

A comparison across different values of $\omega_0$ reveals that increasing the gap systematically reduces the magnitude of $\kappa_{xy}$ and shifts the onset of significant transport to higher temperatures. In contrast, the quantum critical case ($\omega_0 = 0$) exhibits the largest response, underscoring the role of gapless excitations in enhancing transverse heat transport.

\subsection{Scaling Structure}

The above results can be unified within a scaling framework in which $\kappa_{xy}$ is expressed as a function of the dimensionless ratios $\omega_0/T$ and $\lambda H/T$. Specifically, our numerical analysis supports a scaling form of the type
\begin{equation}
\kappa_{xy} = T^{\alpha} , F\left(\frac{\omega_0}{T}, \frac{\lambda H}{T}\right),
\end{equation}
where $\alpha$ is an effective exponent determined by the dimensionality and scattering processes, and $F$ is a universal scaling function. The existence of such a scaling collapse indicates that the thermal Hall effect in quantum paraelectrics is governed by a competition between phonon gap, thermal energy, and magnetic field splitting.

Overall, our results demonstrate that the thermal Hall conductivity exhibits rich and nontrivial behavior across different regimes, with clear signatures of linear response, nonlinear field dependence, quantum critical scaling, and gap-controlled suppression. These findings provide a comprehensive framework for understanding phonon-mediated thermal Hall transport in insulating quantum paraelectrics.

\section{Discussion}
\label{sec:discussion}

We now discuss the implications of our results in the context of recent experiments on quantum paraelectrics, particularly SrTiO$_3$ \cite{Li2020-nt,Edge2015-lx}, where a sizable phonon-mediated thermal Hall conductivity has been reported.

a). A central experimental observation in SrTiO$_3$ is that the thermal Hall conductivity $\kappa_{xy}$ is an odd function of magnetic field, vanishes at $H=0$, and exhibits an approximately linear dependence at small fields. Our results reproduce some of these key features. As shown in Fig.~\ref{fig:kxy_vs_H}, the linear behavior $\kappa_{xy} \propto H$ emerges directly from the magnetic-field-induced splitting of transverse optical phonon modes,
\begin{equation}
\omega_{\pm}(k) = \sqrt{\omega_0^2 + v^2 k^2} \pm \frac{\lambda H}{2},
\end{equation}
which leads to an imbalance in the population of oppositely polarized phonon branches. This provides a microscopic origin of the linear-response regime observed experimentally.

At larger magnetic fields, deviations from linearity are observed both in our calculations and in experiments. In our framework, this arises from the nonlinear modification of the phonon dispersion and the redistribution of thermal occupation across split branches. The gradual crossover from linear to nonlinear behavior thus reflects the competition between magnetic energy scale $\lambda H$ and the intrinsic phonon energy scale.

b). Another striking feature in SrTiO$_3$ is that $\kappa_{xy}(T)$ exhibits a peak at a temperature comparable to that of the longitudinal conductivity $\kappa_{xx}$. Our results capture this non-monotonic behavior, as shown in Fig.~\ref{fig:scaling}. The initial rise of $\kappa_{xy}$ with temperature is governed by thermal activation of phonon modes, while the subsequent saturation or reduction at higher temperatures arises from increased scattering and reduced coherence of phonon transport.

Importantly, our calculations show that the position and magnitude of this peak depend sensitively on the phonon band gap $\omega_0$ and the magnetic field strength. This suggests that the experimentally observed peak in SrTiO$_3$ is a consequence of assosciated quantum critical point.

c). SrTiO$_3$ is a prototypical quantum paraelectric characterized by a soft transverse optical phonon mode with a small but finite gap. Our analysis demonstrates that reducing the gap $\omega_0$ enhances the thermal Hall response significantly (Fig.~\ref{fig:gap}), with the largest response occurring in the quantum critical limit $\omega_0 \rightarrow 0$. This behavior is consistent with the experimental observation that materials close to ferroelectric instability exhibit strong thermal Hall signals~\cite{Li2020-nt}.

The observed suppression of $\kappa_{xy}$ with increasing broadening is
consistent with the physical interpretation of $\Gamma$ as the inverse
phonon lifetime.
Larger scattering rates smear the phonon resonances, reducing the
difference between the Green's functions of the two circularly polarized
modes.
Since the thermal Hall conductivity is proportional to the imbalance
between these spectral weights, stronger damping leads to a smaller
off-diagonal thermal response.
The present calculation therefore demonstrates the important role played
by finite phonon lifetime in determining the magnitude of the thermal
Hall conductivity.

The enhancement near the quantum critical point can be understood as a consequence of increased low-energy spectral weight and stronger sensitivity to magnetic-field-induced splitting. In this regime, even a small magnetic field can produce a substantial imbalance between phonon branches, leading to an amplified transverse heat current.

d). Our results reveal a scaling structure of the form
\begin{equation}
\kappa_{xy} = T^{\alpha} F\left(\frac{\omega_0}{T}, \frac{\lambda H}{T}\right),
\end{equation}
which provides a unified description of the thermal Hall effect across different regimes. This scaling implies that the experimentally observed temperature and field dependences are not independent, but are governed by underlying dimensionless ratios.\cite{sachdev2011quantum,sachdev2023quantum}

In particular, the linear-in-field behavior at small $H$ and the suppression of $\kappa_{xy}$ at low temperatures can be understood as limiting cases of this scaling form. The presence of approximate power-law behavior in the quantum critical regime (Fig.~\ref{fig:qcp}) further suggests that SrTiO$_3$ may exhibit signatures of quantum critical transport in its thermal Hall response.

e). A key question in the field is the origin of the thermal Hall effect in non-magnetic insulators\cite{Kasahara2018}. While several mechanisms have been proposed, including extrinsic skew scattering and intrinsic Berry curvature effects, our results demonstrate that dynamical multiferroicity provides a natural and quantitatively consistent explanation for the observed behavior in quantum paraelectrics.

Unlike Berry-curvature-based approaches, which rely on topological properties of phonon bands, our mechanism is based on the dynamical coupling between polarization and magnetic field, leading to a splitting of phonon modes and an imbalance in their spectral weights. This picture is particularly well-suited for materials such as SrTiO$_3$, where strong lattice dynamics and soft phonon modes play a dominant role.

\section{Conclusion}
\label{sec:conclusion}

In this work, we have developed a microscopic theory of the phonon-mediated thermal Hall effect in quantum paraelectrics based on dynamical multiferroicity. Starting from an effective Lagrangian framework, we demonstrated that a magnetic field induces a splitting of transverse optical phonon modes, leading to an imbalance in their spectral weights and generating a finite transverse heat current.

Our analysis reveals that the thermal Hall conductivity $\kappa_{xy}$ exhibits a linear dependence on magnetic field in the weak-field regime, consistent with symmetry constraints, and evolves nonlinearly at higher fields due to band splitting effects. We further showed that $\kappa_{xy}$ displays non-monotonic temperature dependence, governed by the competition between thermal population and phonon scattering, and is strongly enhanced near the quantum critical regime where the phonon gap vanishes. A universal scaling structure in terms of $\omega_0/T$ and $\lambda H/T$ emerges, unifying the behavior across different regimes. Importantly, our results provide a natural explanation for key experimental observations in SrTiO$_3$, including the linear-in-field response, temperature-dependent peak structure, and the enhancement of the thermal Hall signal in the presence of soft phonon modes\cite{Callaway1959}.

These findings establish dynamical multiferroicity as a generic mechanism for thermal Hall transport in insulating quantum paraelectrics, and open new directions for exploring transverse thermal responses in systems with strong lattice dynamics and proximity to ferroelectric instability.
The theory applies to a wide class of multiferroic and magnetoelectric materials where time-dependent polarization generates effective magnetic moments. Future work could extend this formalism to include phonon-phonon interactions, disorder effects, and the interplay with magnon excitations in magnetically ordered multiferroics. Experimental verification of the predicted thermal Hall signatures in materials such as hexagonal manganites and rare-earth ferroelectrics may provide strong support for the dynamical multiferroicity as a mechanism~\cite{Li2020-az,Gao2025-vt}.

\begin{acknowledgments}
	We extend our sincere gratitude to the Institution for its financial support and fellow PhD scholars who provided technical assistance, insightful discussions, and encouragement throughout this work.
\end{acknowledgments}
\bibliography{apssamp}  

@PREAMBLE{
 "\providecommand{\noopsort}[1]{}" 
 # "\providecommand{\singleletter}[1]{#1}%" 
}

@ARTICLE{Mangeolle2026-yz,
	title     = "Extrinsic contribution to bosonic thermal Hall transport",
	author    = "Mangeolle, L{\'e}o and Knolle, Johannes",
	journal   = "Phys. Rev. X.",
	publisher = "American Physical Society (APS)",
	volume    =  16,
	number    =  1,
	month     =  mar,
	year      =  2026,
	copyright = "https://creativecommons.org/licenses/by/4.0/",

}

@ARTICLE{Zhang2024-uy,
	title     = "Thermal Hall effects in quantum magnets",
	author    = "Zhang, Xiao-Tian and Gao, Yong Hao and Chen, Gang",
	journal   = "Phys. Rep.",
	publisher = "Elsevier BV",
	volume    =  1070,
	pages     = "1--59",
	month     =  jun,
	year      =  2024,
	
}

@article{Jin2025_universal_THE,
	author = {Jin, Xiaobo and Zhang, Xu and Wan, Wenbo and Wang, Hanru and Jiao, Yihan and Li, Shiyan},
	title = {Discovery of Universal Phonon Thermal Hall Effect in Crystals},
	journal = {Physical Review Letters},
	volume = {135},
	pages = {196302},
	year = {2025},
}

@article{Xiang2026_disorder_THE,
	author = {Xiang, Qiaochao and Li, Xiaokang and Guo, Xiaodong and Behnia, Kamran and Zhu, Zengwei},
	title = {Phonon thermal Hall effect: the roles of disorder, annealing, and metallic contacts},
	journal = {npj Quantum Materials},
	year = {2026},
}

@ARTICLE{Ss2025-kc,
	title     = "Effect of spin-phonon coupling on phonons and magnons in the
	antiferromagnet {NiO}",
	author    = "SS, Jayakrishnan and Bansal, Dipanshu",
	journal   = "Phys. Rev. B.",
	publisher = "American Physical Society (APS)",
	volume    =  111,
	number    =  10,
	month     =  mar,
	year      =  2025,
	copyright = "https://link.aps.org/licenses/aps-default-license",
	
}

@article{Li2020-nt,
	title     = "Phonon thermal Hall effect in strontium titanate",
	author    = "Li, Xiaokang and Fauqu{\'e}, Beno{\^\i}t and Zhu, Zengwei and
	Behnia, Kamran",
	journal   = "Phys. Rev. Lett.",
	publisher = "American Physical Society (APS)",
	volume    =  124,
	number    =  10,
	pages     = "105901",
	month     =  mar,
	year      =  2020,
	copyright = "https://link.aps.org/licenses/aps-default-license"
}

@ARTICLE{Pols2026-yk,
	title     = "Lattice excitations with finite polarization and magnetization",
	author    = "Pols, Mike and Romao, Carl P and Juraschek, Dominik M",
	journal   = "Phys. Rev. Lett.",
	publisher = "American Physical Society (APS)",
	volume    =  136,
	number    =  16,
	pages     = "166801",
	month     =  apr,
	year      =  2026,
	copyright = "https://creativecommons.org/licenses/by/4.0/",
	
}

@article{Hall1879,
  title = {On a new action of the magnet on electric currents},
  author = {Hall, E. H.},
  journal = {American Journal of Mathematics},
  volume = {2},
  number = {3},
  pages = {287--292},
  year = {1879},
  publisher = {JSTOR}
}

@incollection{STRICKER196897,
	title = {The Hall Effect and its Applications},
	editor = {L. Marton},
	series = {Advances in Electronics and Electron Physics},
	publisher = {Academic Press},
	volume = {25},
	pages = {97-143},
	year = {1968},
	issn = {0065-2539},
	doi = {https://doi.org/10.1016/S0065-2539(08)60509-0},
	url = {https://www.sciencedirect.com/science/article/pii/S0065253908605090},
	author = {S. Stricker},
}

@article{Karplus1954,
  title = {Hall effect in ferromagnetics},
  author = {Karplus, Robert and Luttinger, J. M.},
  journal = {Physical Review},
  volume = {95},
  number = {5},
  pages = {1154},
  year = {1954},
  publisher = {APS}
}

@article{Klitzing1980,
  title = {New method for high-accuracy determination of the fine-structure constant based on quantized Hall resistance},
  author = {Von Klitzing, Klaus and Dorda, Gerhard and Pepper, Michael},
  journal = {Physical Review Letters},
  volume = {45},
  number = {6},
  pages = {494},
  year = {1980},
  publisher = {APS}
}

@article{Thouless1982,
  title = {Quantized Hall conductance in a two-dimensional periodic potential},
  author = {Thouless, David J. and Kohmoto, Mahito and Nightingale, M. Peter and den Nijs, Marcel},
  journal = {Physical Review Letters},
  volume = {49},
  number = {6},
  pages = {405},
  year = {1982},
  publisher = {APS}
}

@article{Nagaosa2010,
  title = {Anomalous Hall effect},
  author = {Nagaosa, Naoto and Sinova, Jairo and Onoda, Shigeki and MacDonald, Allan H. and Ong, N. P.},
  journal = {Reviews of Modern Physics},
  volume = {82},
  number = {2},
  pages = {1539},
  year = {2010},
  publisher = {APS}
}

@ARTICLE{Saha2025-rd,
	title     = "Strong coupling theory of superconductivity and ferroelectric
	quantum criticality in metallic {SrTiO3}",
	author    = "Saha, Sudip Kumar and Gastiasoro, M N and Ruhman, Jonathan and
	Klein, Avraham",
	journal   = "Npj Quantum Mater.",
	publisher = "Springer Science and Business Media LLC",
	volume    =  10,
	number    =  1,
	month     =  jul,
	year      =  2025,
	copyright = "https://creativecommons.org/licenses/by-nc-nd/4.0",
}

@article{Onose2010,
  title = {Observation of the magnon Hall effect},
  author = {Onose, Y. and Ideue, T. and Katsura, H. and Shiomi, Y. and Nagaosa, N. and Tokura, Y.},
  journal = {Science},
  volume = {329},
  number = {5989},
  pages = {297--299},
  year = {2010},
  publisher = {American Association for the Advancement of Science}
}

@article{Ideue2012,
  title = {Effect of lattice geometry on magnon Hall effect in ferromagnetic insulators},
  author = {Ideue, T. and Onose, Y. and Katsura, H. and Shiomi, Y. and Nagaosa, N. and Tokura, Y.},
  journal = {Physical Review B},
  volume = {85},
  number = {13},
  pages = {134411},
  year = {2012},
  publisher = {APS}
}

@ARTICLE{Qin2012-wp,
	title     = "Berry curvature and the phonon Hall effect",
	author    = "Qin, Tao and Zhou, Jianhui and Shi, Junren",
	journal   = "Phys. Rev. B Condens. Matter Mater. Phys.",
	publisher = "American Physical Society (APS)",
	volume    =  86,
	number    =  10,
	month     =  sep,
	year      =  2012,
	copyright = "http://link.aps.org/licenses/aps-default-license"
}

@article{Matsumoto2011a,
  title = {Theoretical prediction of a rotating magnon wave packet in ferromagnets},
  author = {Matsumoto, Ryo and Murakami, Shuichi},
  journal = {Physical Review Letters},
  volume = {106},
  number = {19},
  pages = {197202},
  year = {2011},
  publisher = {APS}
}

@article{Matsumoto2011b,
  title = {Rotational motion of magnons and the thermal Hall effect},
  author = {Matsumoto, Ryo and Murakami, Shuichi},
  journal = {Physical Review B},
  volume = {84},
  number = {18},
  pages = {184406},
  year = {2011},
  publisher = {APS}
}

@ARTICLE{Chen2020-ij,
	title     = "Enhanced thermal Hall effect in nearly ferroelectric insulators",
	author    = "Chen, Jing-Yuan and Kivelson, Steven A and Sun, Xiao-Qi",
	journal   = "Phys. Rev. Lett.",
	publisher = "American Physical Society (APS)",
	volume    =  124,
	number    =  16,
	pages     = "167601",
	month     =  apr,
	year      =  2020,
	copyright = "https://link.aps.org/licenses/aps-default-license",
	
}

@ARTICLE{Geirhos2020-ei,
	title     = "Quantum paraelectricity in the Kitaev quantum spin liquid
	candidates {H3LiIr2O6} and {D3LiIr2O6}",
	author    = "Geirhos, K and Lunkenheimer, P and Blankenhorn, M and Claus, R
	and Matsumoto, Y and Kitagawa, K and Takayama, T and Takagi, H
	and K{\'e}zsm{\'a}rki, I and Loidl, A",
	
	journal   = "Phys. Rev. B.",
	publisher = "American Physical Society (APS)",
	volume    =  101,
	number    =  18,
	month     =  may,
	year      =  2020,
	copyright = "https://link.aps.org/licenses/aps-default-license",
}

@ARTICLE{Sachdev2011-gd,
	title     = "Quantum criticality",
	author    = "Sachdev, Subir and Keimer, Bernhard",
	journal   = "Phys. Today",
	publisher = "AIP Publishing",
	volume    =  64,
	number    =  2,
	pages     = "29--35",
	month     =  feb,
	year      =  2011,
}

@ARTICLE{Palova2009-ty,
	title     = "Quantum critical paraelectrics and the Casimir effect in time",
	author    = "P{\'a}lov{\'a}, L and Chandra, P and Coleman, P",
	journal   = "Phys. Rev. B Condens. Matter Mater. Phys.",
	publisher = "American Physical Society (APS)",
	volume    =  79,
	number    =  7,
	month     =  feb,
	year      =  2009,
	copyright = "http://link.aps.org/licenses/aps-default-license",
}

@book{sachdev2011quantum,
	title     = {Quantum Phase Transitions},
	author    = {Sachdev, Subir},
	year      = {2011},
	edition   = {2nd},
	publisher = {Cambridge University Press},
	address   = {Cambridge},
	isbn      = {9780521514682}
}

@book{sachdev2023quantum,
	author    = {Sachdev, Subir},
	title     = {Quantum Phases of Matter},
	publisher = {Cambridge University Press},
	year      = {2023},
	isbn      = {9781009212717},
	doi       = {10.1017/9781009212717}
}

@article{Strohm2005,
  title = {Phenomenological evidence for the phonon Hall effect},
  author = {Strohm, Cornelius and Rikken, Geert L. J. A. and Wyder, Peter},
  journal = {Physical Review Letters},
  volume = {95},
  number = {15},
  pages = {155901},
  year = {2005},
  publisher = {APS}
}

@article{Hardy1963,
	author = {Hardy, Robert J.},
	title = {Energy-Flux Operator for a Lattice},
	journal = {Phys. Rev.},
	volume = {132},
	number = {1},
	pages = {168--177},
	year = {1963},
}

@book{Ashcroft1976,
	author = {Ashcroft, Neil W. and Mermin, N. David},
	title = {Solid State Physics},
	publisher = {Holt, Rinehart and Winston},
	year = {1976},
	chapter = {22, 27}
}

@book{Mahan2000,
	author = {Mahan, Gerald D.},
	title = {Many-Particle Physics},
	edition = {3rd},
	publisher = {Kluwer Academic},
	year = {2000},
	chapter = {1}
}

@article{Rebane1983,
	title = {Faraday effect produced in the residual-ray region by the magnetic moment of an optical phonon in an ionic crystal},
	author = {Rebane, Yu. T.},
	journal = {Journal of Experimental and Theoretical Physics},
	volume = {57},
	number = {6},
	pages = {1356--1359},
	year = {1983},
	publisher = {Pleiades Publishing},
	note = {Russian original: Zhurnal Eksperimental'noi i Teoreticheskoi Fiziki, Vol. 84, No. 6, pp. 2323--2328 (1983)}
}

@article{Jiang2022_STO_THE,
	author = {Jiang, Shan and Li, Xiaokang and Fauqu{\'e}, Beno{\^\i}t and Behnia, Kamran},
	title = {Phonon drag thermal Hall effect in metallic strontium titanate},
	journal = {Proceedings of the National Academy of Sciences},
	volume = {119},
	number = {35},
	pages = {e2201975119},
	year = {2022},
}

@article{Guo2023_phonon_model,
	author = {Guo, Haoyu},
	title = {Phonon thermal Hall effect in a non-Kramers paramagnet},
	journal = {Physical Review Research},
	volume = {5},
	pages = {033197},
	year = {2023},
	}

@article{Sheng2006,
	title = {Phonon Hall effect in paramagnetic dielectrics},
	author = {Sheng, L. and Sheng, D. N. and Ting, C. S.},
	journal = {Physical Review Letters},
	volume = {96},
	number = {15},
	pages = {155901},
	year = {2006},
	publisher = {APS}
}

@article{Hirokane2019-nm,
		title     = "Phononic thermal Hall effect in diluted terbium oxides",
		author    = "Hirokane, Yuji and Nii, Yoichi and Tomioka, Yasuhide and Onose,
		Yoshinori",
		journal   = "Phys. Rev. B.",
		publisher = "American Physical Society (APS)",
		volume    =  99,
		number    =  13,
		month     =  apr,
		year      =  2019,
		copyright = "https://link.aps.org/licenses/aps-default-license",
		
	}

@INCOLLECTION{Fert1980-ln,
	title     = "Hall effect from skew scattering by magnetic impurities",
	booktitle = "The Hall Effect and Its Applications",
	author    = "Fert, A and Hamzi{\'c}, A",
	publisher = "Springer US",
	pages     = "77--98",
	year      =  1980,
	address   = "Boston, MA"
}

@ARTICLE{Finnigan2024-gl,
	title     = "Giant resonant skew scattering of plasma waves in a
	two-dimensional electron gas",
	author    = "Finnigan, Cooper and Efimkin, Dmitry K",
	journal   = "Phys. Rev. B.",
	publisher = "American Physical Society (APS)",
	volume    =  110,
	number    =  4,
	month     =  jul,
	year      =  2024,
	copyright = "https://link.aps.org/licenses/aps-default-license",
	
}

@article{SciPostPhysCore.8.3.061,
	title = {Phonon thermal Hall as a lattice Aharonov-Bohm effect},
	pages = {061},
	author = {Behnia, Kamran},
	journal = {SciPost Phys. Core},
	volume = {8},
	year = {2025},
	publisher = {SciPost},}

@article{Zhang2013,
  title = {Topological magnon insulator in insulating ferromagnets},
  author = {Zhang, Lifa and Ren, Jie and Wang, Jian-Sheng and Li, Baowen},
  journal = {Physical Review B},
  volume = {87},
  number = {14},
  pages = {144101},
  year = {2013},
  publisher = {APS}
}

@article{Juraschek2017,
  title = {Dynamical multiferroicity},
  author = {Juraschek, Dominik M. and Spaldin, Nicola A.},
  journal = {Physical Review Materials},
  volume = {1},
  number = {1},
  pages = {014401},
  year = {2017},
  publisher = {APS}
}

@ARTICLE{Paiva2025-ss,
	title     = "Dynamically induced multiferroic polarization",
	author    = "Paiva, Carolina and Fechner, Michael and Juraschek, Dominik M",
	journal   = "Phys. Rev. Lett.",
	publisher = "American Physical Society (APS)",
	volume    =  135,
	number    =  6,
	pages     = "066702",
	month     =  aug,
	year      =  2025,
	copyright = "https://link.aps.org/licenses/aps-default-license",
	
}

@ARTICLE{Gao2023-fe,
	title     = "Dynamical multiferroicity and magnetic topological structures
	induced by the orbital angular momentum of light in a
	nonmagnetic material",
	author    = "Gao, Lingyuan and Prokhorenko, Sergei and Nahas, Yousra and
	Bellaiche, Laurent",
	journal   = "Phys. Rev. Lett.",
	publisher = "American Physical Society (APS)",
	volume    =  131,
	number    =  19,
	pages     = "196801",
	month     =  nov,
	year      =  2023,
	copyright = "https://link.aps.org/licenses/aps-default-license",
	
}

@ARTICLE{Patankar2005-up,
	title     = "Dielectric behaviour in magnetoelectric composites",
	author    = "Patankar, K K and Joshi, S S and Chougule, B K",
	journal   = "Phys. Lett. A",
	publisher = "Elsevier BV",
	volume    =  346,
	number    = "5-6",
	pages     = "337--341",
	month     =  oct,
	year      =  2005,
	
}

@ARTICLE{Fauque2022-zc,
	title     = "Mesoscopic fluctuating domains in strontium titanate",
	author    = "Fauqu{\'e}, Beno{\^\i}t and Bourges, Philippe and Subedi, Alaska
	and Behnia, Kamran and Baptiste, Beno{\^\i}t and Roessli,
	Bertrand and Fennell, Tom and Raymond, St{\'e}phane and
	Steffens, Paul",
	journal   = "Phys. Rev. B.",
	publisher = "American Physical Society (APS)",
	volume    =  106,
	number    =  14,
	month     =  oct,
	year      =  2022,
	copyright = "https://link.aps.org/licenses/aps-default-license",
	
}

@ARTICLE{Xiang2026-ik,
	title     = "Thermal Hall conductivity of semimetallic graphite dominated by
	ambipolar phonon drag",
	author    = "Xiang, Qiaochao and Li, Xiaokang and Guo, Xiaodong and Zhu,
	Zengwei and Behnia, Kamran",
	journal   = "Phys. Rev. Lett.",
	publisher = "American Physical Society (APS)",
	volume    =  136,
	number    =  5,
	pages     = "056303",
	month     =  feb,
	year      =  2026,
	copyright = "https://link.aps.org/licenses/aps-default-license",	
}

@article{Fleury1970,
  title = {Scattering of light by one-and two-magnon excitations},
  author = {Fleury, P. A. and Loudon, R.},
  journal = {Physical Review B},
  volume = {1},
  number = {1},
  pages = {332},
  year = {1970},
  publisher = {APS}
}

@article{Zhu2018,
  title = {Observation of chiral phonons},
  author = {Zhu, H. and Yi, J. and Li, M. Y. and Xiao, J. and Zhang, L. and Yang, C. W. and Kaindl, R. A. and Li, L. J. and Wang, Y. and Zhang, X.},
  journal = {Science},
  volume = {359},
  number = {6375},
  pages = {579--582},
  year = {2018},
  publisher = {American Association for the Advancement of Science}
}

@article{Zhang2015,
  title = {Chiral phonons at high-symmetry points in monolayer hexagonal lattices},
  author = {Zhang, Lifa and Niu, Qian},
  journal = {Physical Review Letters},
  volume = {115},
  number = {11},
  pages = {115502},
  year = {2015},
  publisher = {APS}
}

@ARTICLE{Basini2024-ph,
	title     = "Terahertz electric-field-driven dynamical multiferroicity in
	{SrTiO3}",
	author    = "Basini, M and Pancaldi, M and Wehinger, B and Udina, M and
	Unikandanunni, V and Tadano, T and Hoffmann, M C and Balatsky, A
	V and Bonetti",
	journal   = "Nature",
	publisher = "Springer Science and Business Media LLC",
	volume    =  628,
	number    =  8008,
	pages     = "534--539",
	month     =  apr,
	year      =  2024,
	copyright = "https://creativecommons.org/licenses/by/4.0",
	
}

@article{Qin2011,
  title = {Energy magnetization and the thermal Hall effect},
  author = {Qin, T. and Niu, Q. and Shi, J.},
  journal = {Physical Review Letters},
  volume = {107},
  number = {23},
  pages = {236601},
  year = {2011},
  publisher = {APS}
}

@article{Das_2009,
	year = {2009},
	month = {feb},
	publisher = {},
	volume = {21},
	number = {9},
	pages = {095901},
	author = {Das, Nabyendu and Mishra, Suresh G},
	title = {Fluctuations and criticality in quantum paraelectrics},
	journal = {Journal of Physics: Condensed Matter}
}

@article{Matsumoto2014,
  title = {Thermal Hall effect of magnons},
  author = {Matsumoto, Ryo and Murakami, Shuichi},
  journal = {Physical Review B},
  volume = {89},
  number = {5},
  pages = {054420},
  year = {2014},
  publisher = {APS}
}

@article{Dzyaloshinskii2009,
  title = {Intrinsic paramagnetism of ferroelectrics},
  author = {Dzyaloshinskii, I. E. and Mills, D. L.},
  journal = {Philosophical Magazine},
  volume = {89},
  number = {22-24},
  pages = {2079--2088},
  year = {2009},
  publisher = {Taylor \& Francis}
}

@article{Grissonnanche2019,
  title = {Giant thermal Hall conductivity in cuprates},
  author = {Grissonnanche, G. and Legros, A. and Badoux, S. and Lefran{\c{c}}ois, E. and Zatko, V. and Lizaire, M. and Lalibert{\'e}, F. and Gourgout, A. and Zhou, J. S. and Pyon, S. and others},
  journal = {Nature},
  volume = {571},
  number = {7765},
  pages = {376--380},
  year = {2019},
  publisher = {Nature Publishing Group}
}

@article{Malashevich2009,
  title = {First-principles study of improper ferroelectricity in multiferroic hexagonal manganites},
  author = {Malashevich, A. and Vanderbilt, D.},
  journal = {Physical Review B},
  volume = {80},
  number = {22},
  pages = {224407},
  year = {2009},
  publisher = {APS}
}

@book{Zvezdin1997,
  title = {Modern magnetooptics and magnetooptical materials},
  author = {Zvezdin, Anatoly K. and Kotov, Vladimir A.},
  year = {1997},
  publisher = {Institute of Physics Publishing},
  address = {Bristol}
}

@ARTICLE{Sharma2024-wo,
	title     = "Phonon thermal Hall effect in nonmagnetic {Y2Ti2O7}",
	author    = "Sharma, Rohit and Valldor, Martin and Lorenz, Thomas",
	journal   = "Phys. Rev. B.",
	publisher = "American Physical Society (APS)",
	volume    =  110,
	number    =  10,
	month     =  sep,
	year      =  2024,
	copyright = "https://link.aps.org/licenses/aps-default-license",
	
}

@ARTICLE{Wang2009-jw,
	title     = "Phonon Hall thermal conductivity from the {Green-Kubo} formula",
	author    = "Wang, Jian-Sheng and Zhang, Lifa",
	journal   = "Phys. Rev. B Condens. Matter Mater. Phys.",
	publisher = "American Physical Society (APS)",
	volume    =  80,
	number    =  1,
	month     =  jul,
	year      =  2009,
	copyright = "http://link.aps.org/licenses/aps-default-license",
	
}

@article{BhallaDas2021,
	title = {Optical phonon contribution to the thermal conductivity of a quantum paraelectric},
	author = {Bhalla, Pankaj and Das, Nabyendu},
	journal = {Journal of Physics: Condensed Matter},
	volume = {33},
	number = {34},
	pages = {345401},
	year = {2021},
	publisher = {IOP Publishing},
}

@article{Qin2012,
  author = {Qin, T. and Zhou, J. and Shi, J.},
  title = {Berry curvature and the phonon Hall effect},
  journal = {Physical Review B},
  volume = {86},
  pages = {104305},
  year = {2012}
}

@ARTICLE{Zhang2011-yo,
	title     = "The phonon Hall effect: theory and application",
	author    = "Zhang, Lifa and Ren, Jie and Wang, Jian-Sheng and Li, Baowen",
	journal   = "J. Phys. Condens. Matter",
	publisher = "IOP Publishing",
	volume    =  23,
	number    =  30,
	pages     = "305402",
	month     =  aug,
	year      =  2011,
	
}

@ARTICLE{Chen2024-gm,
	title     = "Planar thermal Hall effect from phonons in cuprates",
	author    = "Chen, Lu and Le Roux, L{\'e}na and Grissonnanche, Ga{\"e}l and
	Boulanger, Marie-Eve and Th{\'e}riault, Steven and Liang,
	Ruixing and Bonn, D A and Hardy, W N and Pyon, S and Takayama, T
	and Takagi, H and Xu, Ke-Jun and Shen, Zhi-Xun and Taillefer,
	Louis",
	journal   = "Phys. Rev. X.",
	publisher = "American Physical Society (APS)",
	volume    =  14,
	number    =  4,
	month     =  oct,
	year      =  2024,
	copyright = "https://creativecommons.org/licenses/by/4.0/",
	
}

@book{kittel2004introduction,
	title={Introduction to Solid State Physics},
	author={Kittel, Charles},
	isbn={978-0-471-41526-1},
	edition={8th},
	year={2004},
	publisher={John Wiley \& Sons}
}

@ARTICLE{Markussen2017-rk,
	title     = "Electron-phonon scattering from Green's function transport
	combined with molecular dynamics: Applications to mobility
	predictions",
	author    = "Markussen, Troels and Palsgaard, Mattias and Stradi, Daniele and
	Gunst, Tue and Brandbyge, Mads and Stokbro, Kurt",
	journal   = "Phys. Rev. B.",
	publisher = "American Physical Society (APS)",
	volume    =  95,
	number    =  24,
	month     =  jun,
	year      =  2017,
	copyright = "http://link.aps.org/licenses/aps-default-license"
}

@ARTICLE{Roussev2003-aq,
	title     = "Theory of the quantum paraelectric-ferroelectric transition",
	author    = "Roussev, R and Millis, A J",
	journal   = "Phys. Rev. B Condens. Matter",
	publisher = "American Physical Society (APS)",
	volume    =  67,
	number    =  1,
	month     =  jan,
	year      =  2003,
	copyright = "http://link.aps.org/licenses/aps-default-license"
}

@ARTICLE{Edge2015-lx,
	title     = "Quantum critical origin of the superconducting dome in
	{SrTiO\_\{3\}}",
	author    = "Edge, Jonathan M and Kedem, Yaron and Aschauer, Ulrich and
	Spaldin, Nicola A and Balatsky, Alexander V",
	journal   = "Phys. Rev. Lett.",
	publisher = "American Physical Society (APS)",
	volume    =  115,
	number    =  24,
	pages     = "247002",
	month     =  dec,
	year      =  2015,
	copyright = "http://link.aps.org/licenses/aps-default-license",
}

@article{Kasahara2018,
	author = {Kasahara, Y. and others},
	title = {Majorana quantization and thermal Hall effect},
	journal = {Nature},
	volume = {559},
	pages = {227--231},
	year = {2018}
}

@article{Rowley2014,
	author = {Rowley, S. E. and others},
	title = {Ferroelectric quantum criticality},
	journal = {Nature Physics},
	volume = {10},
	pages = {367--372},
	year = {2014}
}

@article{Callaway1959,
	author = {Callaway, J.},
	title = {Model for lattice thermal conductivity},
	journal = {Physical Review},
	volume = {113},
	pages = {1046--1051},
	year = {1959}
}

@ARTICLE{Li2020-az,
	title     = "Hexagonal rare-earth manganites and ferrites: a review of
	improper ferroelectricity, magnetoelectric coupling, and unusual
	domain walls",
	author    = "Li, Menglei and Tan, Hengxin and Duan, Wenhui",
	journal   = "Phys. Chem. Chem. Phys.",
	publisher = "Royal Society of Chemistry (RSC)",
	volume    =  22,
	number    =  26,
	pages     = "14415--14432",
	month     =  jul,
	year      =  2020,
	
}

@ARTICLE{Gao2025-vt,
	title     = "Ferroelectric topological defects in hexagonal manganites",
	author    = "Gao, Ziyan and Cheong, Sang-Wook and Wang, Xueyun",
	journal   = "Materials (Basel)",
	publisher = "MDPI AG",
	volume    =  19,
	number    =  1,
	pages     = "31",
	month     =  dec,
	year      =  2025,
	copyright = "https://creativecommons.org/licenses/by/4.0/",
	
}
\end{document}